\pdfoutput=1
\documentclass[final,5p,margin=1in,nopreprintline]{elsarticle}

\usepackage{amsmath,amsfonts,amssymb}
\usepackage{graphicx}
\usepackage{hyperref}
\usepackage{graphicx}
\usepackage{xcolor}
\usepackage{booktabs}
\usepackage{enumitem}
\usepackage{algorithm,algpseudocode}
\usepackage{adjustbox}
\usepackage{booktabs}
\usepackage{makecell}

\newcommand{\model}{\theta}
\algrenewcommand{\algorithmiccomment}[1]{\hfill$\triangleright$~#1}

\begin{document}

\begin{frontmatter}
\title{Robust and Private Swarm Learning via Zero-Knowledge Validator Admission and Consensus Aggregation}

\author[1]{Marco Arazzi\fnref{alpha}}
\ead{marco.arazzi01@universitadipavia.it}

\author[1]{Mert Cihangiroglu\corref{cor1}\fnref{alpha}}
\ead{mert.cihangiroglu01@universitadipavia.it}

\author[1]{Antonino Nocera\fnref{alpha}}
\ead{antonino.nocera@unipv.it}

\cortext[cor1]{Corresponding author.}
\fntext[alpha]{Authors are listed in alphabetical order.}

\affiliation[1]{organization={Department of Electrical, Computer and Biomedical Engineering, University~of~Pavia},
            addressline={Via~A.~Ferrata,~5},
            city={Pavia},
            postcode={27100},
            state={PV},
            country={Italy}}

\begin{abstract}
Swarm Learning improves resilience and eliminates the need for a single trusted coordinator by distributing model aggregation and validation across participating nodes. However, it also delegates the validation of model updates to peer nodes whose private data may be biased, unrepresentative, or adversarial. Existing Swarm Learning protocols generally do not attest the data used by validators, leaving decentralized validation as an attack surface. We present a robust, privacy-preserving Swarm Learning framework that makes verifiable data quality the condition for admission to the learning task. Specifically, a validator joins only after a zero-knowledge attestation proves, without revealing any raw sample, that a reference model ran correctly on its private data. The attested outputs also yield a label-free class profile, so the network assembles a collectively IID validator pool from individually skewed validators as accurately as an oracle with the true labels (total-variation distance to uniform 0.025, against 0.021 for the oracle and 0.21 for random selection). Admitted validators score client updates and, through a Tendermint-style Byzantine Fault Tolerant consensus, agree on a gradient mask and a quality-derived aggregation weight for each submission, replacing dataset-size weighting with a privacy-preserving, quality-aware signal. Adapting the cut-layer idea of Split Learning, a challenge-based audit lets peers re-verify that a mask was computed on the attested data without an expensive end-to-end zero-knowledge proof. Across four benchmarks under non-IID partitions the framework matches FedAvg on clean data; under a gradient-scaling attack that collapses FedAvg to random chance, it retains 94\% of clean accuracy on CIFAR-100 and over 99\% on FashionMNIST, where FedAvg achieves only 2.1\% and 11\%. The result is a private, robust, and auditable foundation for decentralized learning.
\end{abstract}

\begin{keyword}
Swarm Learning, Federated Learning, Byzantine Fault Tolerance, Non-IID Data, Zero-Knowledge Proofs, Gradient Masking, Blockchain, Privacy Preservation
\end{keyword}

\end{frontmatter}

\section{Introduction}
Swarm Learning (SL) trains a shared model across many data holders without a central server and without moving raw data off site \cite{warnat2021swarm}. This decentralization makes it attractive for healthcare, finance, and cybersecurity, where data is sensitive and direct sharing is often legally restricted. Removing the server also removes the one party that Federated Learning (FL) relies on to evaluate whether a model update has improved \cite{mcmahan2017communication}. SL transfers this task to \emph{validators}, which use their own private data to validate candidate models and decide how each contribution is aggregated. Every aggregation decision therefore depends on the quality of the validators' data, yet no existing SL protocol checks that quality. A validator may hold data that is skewed, stale, or adversarial, and the SL network has no way to verify this situation.

Unchecked validator data is indeed a security issue, not just a statistical problem. An adversary admitted as a validator can steer the aggregation with a few well-placed scores \cite{fang2020local}, and a Sybil adversary can register many such validators at little cost. Byzantine-robust aggregation rules such as Krum, Trimmed-Mean, Bulyan, and Flame \cite{blanchard2017machine,yin2018byzantine,guerraoui2018hidden,nguyen2022flame} harden the aggregation step by treating suspicious updates as statistical outliers, but assume a trusted aggregator and never verify the data behind a participant's vote. Methods that target heterogeneous, non-IID data, such as FedProx, FedNova, and SCAFFOLD \cite{li2020federated,wang2020tackling,karimireddy2020scaffold}, correct for client drift but assume honest participants throughout. None of them certify that the data used to evaluate and weight a model is representative and genuine, so in a decentralized setting validators remain an unguarded attack surface.

In our proposal, we attempt to address this gap with a robust and privacy-preserving Swarm Learning framework whose validators are admitted, audited, and weighted by a \emph{verifiable}, privacy-preserving test on their private data. The framework joins three mechanisms under one Byzantine Fault Tolerant (BFT) workflow inspired by Tendermint \cite{buchman2016tendermint}: a zero-knowledge attestation that gates validator admission on proven data quality \cite{goldwasser2019knowledge,halo2}, a quality-aware mask consensus that sets each aggregation weight from validator scores rather than from dataset size, and a challenge-based audit adapted from Split Learning \cite{vepakomma2018split} that re-verifies a validator's work without an expensive proof over the full training pass. Each mechanism addresses a specific flaw introduced by the removal of the central server.

The payoff is robustness under attack. A gradient-scaling attack collapses FedAvg to random-chance accuracy, while our framework retains most of its clean accuracy at the same attacker fractions and degrades gradually rather than collapsing. On clean data, the framework matches FedAvg overall, with its gains concentrated on the most heterogeneous partitions, where each client sees only a narrow slice of the label space. Our contributions are as follows.

\begin{itemize}
    \item We introduce an admission protocol in which a candidate cryptographically proves that a reference model ran correctly on its committed private dataset, without revealing raw inputs. Additionally, admission requires a confidence-based score to exceed a specified threshold. We assume authenticated membership in a permissioned deployment; identity uniqueness and prevention of dataset reuse are outside the admission protocol. The same attested outputs yield a label-free class profile that lets the network compose a collectively IID validator pool from individually skewed validators, as accurately as an oracle with the true labels.
    \item Admitted validators score client models on their private data and, through a BFT consensus inspired by Tendermint, agree on a gradient-based mask and a consensus-derived aggregation weight for each submission, replacing dataset-size weighting with a privacy-preserving signal that down-weights low-quality and adversarial contributions.
    \item Adapting the cut-layer principle of Split Learning, we design an audit that lets peers confirm that a validator's mask was genuinely computed on its attested data, deterring stealthy poisoning without a prohibitively large zero-knowledge circuit over the full backward pass.
    \item Under non-IID partitions we show clean-data accuracy that matches FedAvg overall and improves most under high heterogeneity, near-complete robustness under gradient-scaling attacks that drive FedAvg to random chance, and ablations over data heterogeneity and validator-quorum size.
\end{itemize}

The remainder of the paper is organized as follows. Section~\ref{sec:prelims} reviews the necessary background; Section~\ref{sec:threat_model} defines the threat model; Section~\ref{sec:description_of_approach} details the proposed framework and its components; Section~\ref{sec:setup} describes the experimental setup; Section~\ref{sec:experiments} reports the results; Section~\ref{sec:limitations} examines scalability and limitations; Section~\ref{sec:related} surveys related work; and Section~\ref{sec:conclusion} presents conclusions and future directions.

\section{Preliminaries}
\label{sec:prelims}
This section defines the four building blocks of our framework and fixes the notation used later: Federated Learning, its decentralized successor Swarm Learning, zero-knowledge proofs, and the blockchain consensus layer that coordinates them. This section also characterizes the non IID data heterogeneity that complicates decentralized validation, which is the challenge our verifiable admission mechanism is designed to address.

\subsection{Federated Learning}

Federated Learning (FL) \cite{mcmahan2017communication,konevcny2016federated,bonawitz2017practical,bonawitz2019towards,li2020challenges} enables multiple distributed clients to collaboratively train a shared global model under the coordination of a central server without exchanging raw data. Each client performs local training on its private dataset and communicates only model updates (e.g., gradients or model parameters) to the server, which aggregates them to update the global model. Although this setting preserves data locality and improves privacy, it also introduces several key challenges.

A first limitation is the presence of non-IID data across clients, where local datasets follow heterogeneous distributions. This typically leads to client drift and can slow the convergence of the global model. A second issue is communication overhead, since model updates must be exchanged between clients and server at every round, which becomes costly in large-scale settings. Finally, validation is usually centralized on the server, creating a bottleneck, as well as a single point of failure and a potential security risk.

Formally, FL is commonly categorized into Horizontal Federated Learning (HFL), Vertical Federated Learning (VFL), and Federated Transfer Learning (FTL) \cite{yang2019federated}. In HFL, clients hold datasets that share the same feature space but correspond to different samples. Each client trains a local model on its private data and sends its parameters $\{\model^i \mid i \in n\}$ to the central server for aggregation.

The HFL objective is defined as:
\[
\min_{\model} \ \ell(\model) = \sum_{i=1}^{n} \frac{k_i}{n} L_i(\model),
\quad
L_i(\model) = \frac{1}{k_i} \sum_{j \in P_i} \ell_j(\model, x_j),
\]
where $L_i(\model)$ denotes the local loss of client $i$, $k_i$ is the number of local samples, and $P_i$ is the set of indexes of its data points.

In the $m$-th communication round, the training proceeds by first broadcasting the global model $\model_m$ to all clients. Each client then performs several steps of local optimization over its private dataset:
\[
\model_m^i \leftarrow \model_m^i - \alpha \frac{\partial L(\model_m, b)}{\partial \model_m^i},
\]
where $\alpha$ is the learning rate and $b$ denotes a mini-batch. The updated local models are then sent back to the server, which aggregates them to obtain the next global model.

\subsection{Swarm Learning}

Swarm Learning (SL) \cite{warnat2021swarm} extends Federated Learning by eliminating the central aggregator and distributing training and validation across a peer-to-peer network. Coordination can be achieved through blockchain-based smart contracts and consensus mechanisms rather than a central server. As a result, no single participant maintains the global model or controls how updates are combined.

In this setting, model updates are exchanged and merged through a decentralized aggregation rule of the form:
\[
\model_{t+1} = \mathcal{C}\big(\{\model_t^i\}_{i=1}^{n}\big),
\]
where $\mathcal{C}(\cdot)$ denotes a consensus operator over the participating nodes.

This decentralization particularly affects the validation process. In contrast to FL, in which a central server evaluates model performance, SL assigns this role to a set of \textit{validators}, i.e., peers that independently assess candidate models using their local data. Each validator produces a local score:
\[
s_i = \ell(\model, \mathcal{D}_i^{val}),
\]
which reflects the model quality on its private validation data.

Although this increases robustness against central failure, it also introduces several challenges that motivate our framework. First, validators may behave maliciously and return biased or adversarial scores, corrupting the evaluation process. Second, validator data is not guaranteed to be high-quality or representative, which can lead to unreliable assessments. Third, since validation is distributed, a consensus mechanism is required to aggregate individual evaluations into a single decision, which is inherently non-trivial in fully decentralized settings.

\subsection{Zero-Knowledge Proofs}

A zero-knowledge proof (ZKP) lets a prover convince a verifier that a statement is true, while revealing nothing beyond its validity \cite{goldwasser2019knowledge}. ZKPs support privacy-preserving systems such as blockchain transactions, identity verification, and verifiable computation. We use a ZKP at the admission of each validator. In particular, a candidate proves that it executed the agreed reference model correctly on its committed private test set, binding the input, the model, and the outputs together without disclosing a single raw sample. The reference model and the output distribution are public; only the raw input remains hidden. The proof certifies \emph{how} the computation was run, and the resulting output distributions then feed the data-quality score of Section~\ref{subsec:data_quality_assessment} that decides admission. This keeps decentralized validation both verifiable and privacy-preserving, while guaranteeing that the data of the validators are never exposed.

\begin{table}[ht!]
\centering
\caption{Key Acronyms and Their Full Terms}
\label{tab:acronyms_short}
\begin{tabular}{ll}
\toprule
\textbf{Acronym} & \textbf{Full Term} \\
\midrule
FL   & Federated Learning \cite{mcmahan2017communication} \\
SL   & Swarm Learning  \cite{warnat2021swarm} \\
ZKP  & Zero-Knowledge Proof \cite{goldwasser2019knowledge} \\
BFT  & Byzantine Fault Tolerant \cite{castro1999practical,buchman2016tendermint} \\
MPC  & Multi-Party Computation \cite{yao1982protocols} \\
R1CS & Rank-1 Constraint System \cite{groth2016size} \\
EVM  & Ethereum Virtual Machine \cite{wood2014ethereum} \\
KZG  & Kate-Zaverucha-Goldberg \cite{kate2010constant} \\
SRS  & Structured Reference String \\
ONNX & Open Neural Network Exchange \\

\bottomrule
\end{tabular}
\end{table}

\subsection{Blockchain and Tendermint Consensus}
A blockchain is a replicated append-only ledger maintained by a peer-to-peer network, in which committed transactions are ordered, immutable, and independently verifiable~\cite{wood2014ethereum}. Programmable \emph{smart contracts} let application logic, such as registering model references or verifying submitted proofs, execute deterministically against this shared state. Because no single party controls the ledger, it gives Swarm Learning the decentralized, tamper-evident coordination layer it needs. Agreement on the next block is reached through a consensus protocol. We build on Tendermint~\cite{buchman2016tendermint}, a BFT protocol in the lineage of PBFT~\cite{castro1999practical} that finalizes a value once more than two-thirds of the participants vote for it. A network of $3f+1$ validators tolerates up to $f$ Byzantine faults. A proposer broadcasts a candidate, validators exchange pre-vote and pre-commit messages, and the value commits once it gathers $2f+1$ matching pre-commits. We adapt this quorum structure in Section~\ref{subsec:bft_consensus} to finalize gradient masks rather than financial transactions.

\section{Threat Model}
\label{sec:threat_model}

We consider a decentralized Swarm Learning deployment in which mutually distrusting participants collaboratively train a shared model without a central server. We first outline the trust assumptions on the system components and then describe the adversarial model targeted by our framework.

\textbf{Trusted components.} We assume correctness of the underlying blockchain and its smart contracts: transactions are executed faithfully, the committed state is immutable, and the \texttt{Coordinator} and \texttt{Verifier} contracts correctly store references and verify submitted proofs. Cryptographic primitives are assumed to be secure: the HALO2 proof system is sound and zero-knowledge, while the Poseidon hash~\cite{grassi2021poseidon} and KZG commitments~\cite{kate2010constant} are collision-resistant and binding. We further assume an honest setup phase for generating the structured reference string (SRS). Off-chain storage (IPFS) is trusted only for availability; integrity is ensured via on-chain content hashes, making any tampering detectable.

\textbf{Adversarial participants.} Both clients and validators may act in a Byzantine manner. Malicious clients aim to corrupt the global model by sending poisoned updates, while malicious validators attempt to bias the evaluation by providing unrepresentative data, fabricating gradient masks, or manipulating consensus scores. Following a standard Byzantine fault model, we assume that at most $f$ of the $3f+1$ validators are Byzantine, ensuring that honest validators maintain a $2f+1$ majority required for finalization (Section~\ref{subsec:bft_consensus}). Adversaries may collude and share internal state, but remain computationally bounded and cannot forge proofs or break the underlying cryptographic assumptions.

\textbf{Adversary goals.} We consider three main objectives: (i) \emph{convergence sabotage}, i.e., degrading or preventing convergence of the global model; (ii) \emph{evaluation gaming}, i.e., gaining or maintaining validator influence using unrepresentative, biased, or fabricated data in order to distort the evaluation of the model; and (iii) \emph{Sybil inflation}, i.e., creating multiple identities to increase influence over consensus.

\textbf{Adversary capabilities.} Poisoning adversaries are assumed to be \emph{client-local}, meaning each attacker can only manipulate its own data and submitted updates and observes only the latest published global model. Formally, the attacker view in round $t$ is restricted to
\[
\mathcal{V}_t^{adv} = \{\model_t\},
\]
with no access to concurrent honest updates before aggregation.

Honest-but-curious participants can inspect publicly available artifacts, including cryptographic commitments, gradient masks, and intermediate protocol outputs, and can attempt to infer private information from them. In particular, a validator’s published output distribution
\[
p_i(y) = \frac{1}{|\mathcal{D}_i|} \sum_{(x,y)\in \mathcal{D}_i} \mathbb{I}[y],
\]
reveals the class composition of its dataset without exposing individual samples.

\textbf{Scope.} The mechanisms addressing these threats—\allowbreak zero-knowledge validator admission, challenge-based mask auditing, and BFT consensus, are detailed in Section~\ref{sec:description_of_approach}. Network-level attacks (e.g., partitioning or eclipse) and attacks on the underlying blockchain consensus layer are orthogonal to our contribution and are considered out of scope.

\section{Description of Our Approach}
\label{sec:description_of_approach}

\begin{figure*}[t]
    \centering
    \includegraphics[width=0.82\linewidth]{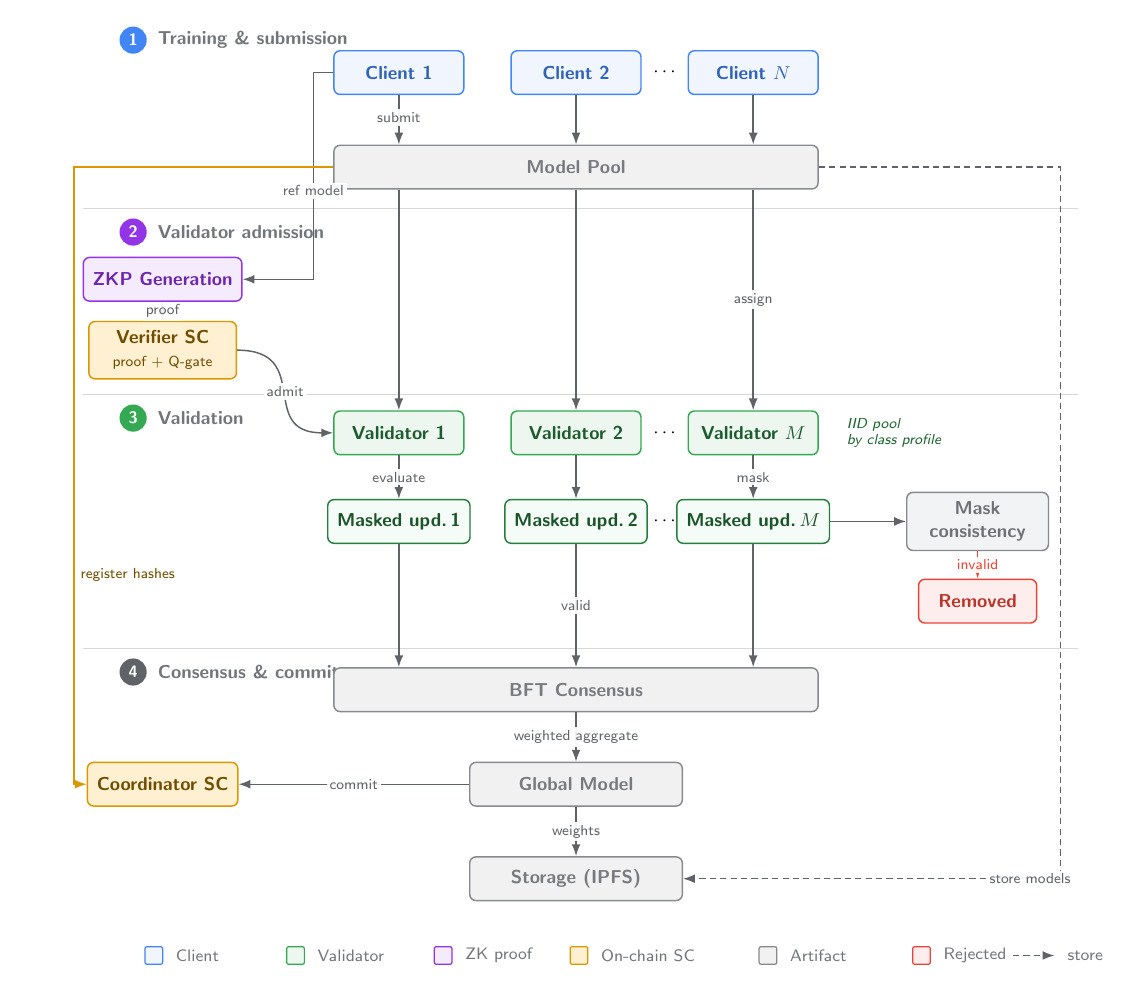}
    \caption{End-to-end architecture of the robust and private Swarm Learning framework, shown across the four numbered stages of one round. \textbf{(1)}~Clients train local models and send them to the Model Pool. Each model is also uploaded to IPFS, and its hash is registered on-chain with the Coordinator SC, so every submission is authenticated and cannot be altered. \textbf{(2)}~A client seeking validator status runs the reference model on its private data and sends the on-chain Verifier SC a zero-knowledge proof that it ran the model correctly. The Verifier SC verifies the proof, and the client is admitted only if its outputs also pass a quality gate that the network applies off-chain, after which the mix of classes the client's data contains (its class profile) is recorded. The admitted validators are picked so that, together, their data is balanced (IID) across all classes. \textbf{(3)}~Each validator scores the pooled client models and builds a gradient mask for every one. A consistency check then removes any validator whose mask its peers cannot reproduce. \textbf{(4)}~The validators run a BFT consensus to agree on the masks and combine the client updates using weights set by that consensus. The result is the next global model, which is committed on-chain and stored on IPFS.}
    \label{fig:system_architecture}
\end{figure*}

Federated Learning and Swarm Learning preserve the privacy of data by not sharing it, but aggregation or validation still relies on trusted entities that may become single points of failure. To address this limitation, we propose a blockchain-based Swarm Learning framework that decentralizes model validation across a set of validator nodes. Validators are admitted through a zero-knowledge attestation protocol that proves correct execution of a reference model while preserving data privacy. Once admitted, they evaluate client models, generate gradient-based masks following \cite{arazzi2025privacy}, and participate in Byzantine Fault Tolerant (BFT) consensus to finalize the global model. Figure~\ref{fig:system_architecture} provides an overview of the framework. Figure~\ref{fig:system_architecture} gives a general overview of the architecture. The framework comprises the following entities:

\begin{itemize}
    \item \textbf{Swarm Clients.} Nodes holding private datasets that train local models and submit them to the model pool.
    
    \item \textbf{Validator Nodes.} Clients admitted through zero-knowledge attestation that evaluate models, generate gradient masks, participate in BFT consensus, and undergo periodic audits.

    \item \textbf{Model Pool.} A decentralized buffer that collects client-submitted models during each training round. Validators asynchronously pull from this pool to perform evaluations and compute class-specific gradient masks.

    \item \textbf{Blockchain Layer.} Coordinates the protocol through smart contracts, stores model references and validator metadata, and verifies zero-knowledge proofs.

    \item \textbf{IPFS or Off-Chain Storage.} Used to store complete versions of local and global models, activations, or logs, referenced through cryptographic hashes in blockchain transactions. This decouples data availability from consensus weight.
\end{itemize}

The lifecycle of a Swarm Learning round in our framework consists of the following phases, which are also shown in Figure~\ref{fig:system_architecture}:

\begin{enumerate}
    \item \textbf{Local Training.} Clients train local models, upload them to IPFS, and register their hashes on-chain.
    
    \item \textbf{Evaluation and Masking.} Validators evaluate submitted models on private validation data and compute gradient-based masks.
    
    \item \textbf{Consensus.} Validators exchange evaluations and finalize the global model through BFT consensus.
    
    \item \textbf{Audit.} Selected validators undergo challenge-based verification to confirm the correctness of their published masks.
\end{enumerate}

The following subsections describe each component of the framework: the system model, the validator admission protocol, gradient-mask generation, challenge-based verification, and BFT consensus.

\subsection{System Model}
\label{subsec:system_model}

Our framework involves a finite set of \mbox{$N$} participants, \mbox{$\mathcal{N} = \{ n_1, n_2, \dots, n_N \}$}. Each participant \mbox{$n_i \in \mathcal{N}$} holds a private local dataset:
\begin{equation}
\mathcal{D}^{(i)} = \left\{ (\mathbf{x}_j^{(i)}, y_j^{(i)}) \right\}_{j=1}^{|\mathcal{D}^{(i)}|},
\end{equation}
where \mbox{$\mathbf{x}_j^{(i)} \in \mathbb{R}^{d}$} is the \mbox{$j$}-th input held by participant \mbox{$n_i$} and \mbox{$y_j^{(i)} \in \mathcal{Y}$} its label. Each dataset \mbox{$\mathcal{D}^{(i)}$} is drawn from a local distribution \mbox{$\mathcal{P}_i(\mathbf{x}, y)$}, and these distributions differ between participants: for any two participants \mbox{$n_a, n_b \in \mathcal{N}$}, in general \mbox{$\mathcal{P}_a \neq \mathcal{P}_b$}. This is the non-IID setting our design targets. Every local dataset stays private, so no participant shares or exposes its raw samples. Participants jointly train a global model with weights \mbox{$\mathbf{w} \in \mathbb{R}^m$} by minimizing a weighted sum of their local losses. FedAvg weights each participant by its relative dataset size:
\begin{equation}
\mathcal{L}(\mathbf{w}) = \sum_{i=1}^{N} \frac{|\mathcal{D}^{(i)}|}{\sum_{k=1}^{N}|\mathcal{D}^{(k)}|} \mathcal{L}^{(i)}(\mathbf{w}),
\end{equation}
where the local loss $\mathcal{L}^{(i)}(\mathbf{w})$ for the participant $n_i$ is given by:
\begin{equation}
\mathcal{L}^{(i)}(\mathbf{w}) = \frac{1}{|\mathcal{D}^{(i)}|} \sum_{j=1}^{|\mathcal{D}^{(i)}|} \ell\left(f(\mathbf{x}_j^{(i)}; \mathbf{w}), y_j^{(i)}\right).
\end{equation}
Here, \mbox{$f(\mathbf{x};\mathbf{w}): \mathbb{R}^{d} \rightarrow \mathcal{Y}$} is the model's prediction and \mbox{$\ell(\cdot)$} a differentiable loss such as cross-entropy.

A subset of the participants act as validators, \mbox{$\mathcal{V} \subseteq \mathcal{N}$}. A participant joins this set only after its dataset passes a verifiable quality check \mbox{$\text{ValidatorSelection}(\cdot)$}, detailed in Section~\mbox{\ref{subsec:validator_selection_zkp}}. In each training round \mbox{$t$}, the validators then perform two additional functions. They turn the pooled client models into gradient masks through \mbox{$\text{GradientMasking}(\cdot)$} (Section~\mbox{\ref{subsec:gradient_mask_generation}}), and they finalize the next global model through a Byzantine Fault Tolerant procedure \mbox{$\text{BFTConsensus}(\cdot)$} (Section~\mbox{\ref{subsec:bft_consensus}}) that commits a candidate only with supermajority support.

The weighted-risk objective above is the conventional FedAvg formulation, shown as a reference. Our framework keeps this structure, but does \emph{not} reuse its static dataset-size weights at aggregation time. Instead, each client's aggregation weight comes from the validator consensus scores of Eq.~\mbox{\ref{eq:consensus-weights}}, so no client has to report its dataset size; Section~\mbox{\ref{sec:weight-evolution}} analyzes how these weights behave in practice. The three functions give the framework its three properties: verifiable admission, local data privacy, and resilience to adversarial updates. Verifiable admission and privacy follow from the zero-knowledge attestation and commitments of Section~\mbox{\ref{subsec:validator_selection_zkp}}. Resilience is not a formal guarantee; we demonstrate it empirically against gradient-scaling attacks in Section~\mbox{\ref{sub:scaling_robustness}}.

\subsection{Verifiable Inference for Validator Admission}
\label{subsec:validator_selection_zkp}

Validator selection is critical to the integrity and robustness of our Swarm Learning framework. To ensure that validators hold representative datasets without compromising their privacy, we integrate a ZKP-based attestation protocol using the EZKL Python library \cite{ezkl}. The following structured protocol governs the validator admission.

\vspace{0.5em}
\textit{Setup Phase.}
The network performs a one-time setup that compiles the reference model into an EZKL circuit, generates the SRS and the proving/verification keys $(pk,vk)$, stores the SRS on IPFS, and deploys the on-chain \texttt{Verifier SC}.

\vspace{0.5em}
\textit{Step 1: Local Inference.} Validator $v_i$ executes the reference model $\mathcal{M}_{\text{ref}}$ on its private test dataset $\mathcal{D}^{(i)}_{\text{test}} = \{x_1, x_2, \dots, x_{n_i}\}$, yielding the output vector:
\begin{equation}
\mathbf{y}^{(i)} = \left(\mathcal{M}_{\text{ref}}(x_1), \dots, \mathcal{M}_{\text{ref}}(x_{n_i})\right).
\end{equation}

\vspace{0.5em}
\textit{Step 2: Commitment Creation.} Before generating a ZKP, $v_i$ creates cryptographic commitments:
\begin{itemize}
    \item \textbf{Input Commitment:} $C(PH(x)) = g^{PH(x)}$ where $PH(x)$ is the Poseidon hash~\cite{grassi2021poseidon} of the dataset.
    \item \textbf{Model Commitment:} $C(PH(M)) = g^{PH(M)}$ for the ONNX reference model.
    \item \textbf{Public Output Commitment:} $C(y) = g^y$ binds the revealed output vector to the proof.
\end{itemize}
The commitments bind the dataset, the reference model, and the public outputs to the proof. Poseidon hashes~\cite{grassi2021poseidon} and KZG commitments~\cite{kate2010constant} ensure that these values cannot be modified after proof generation, while keeping the private dataset hidden.

\vspace{0.5em}
\textit{Step 3: Proof Generation.} Validator $v_i$ uses the proving key and the constructed commitments to generate a proof $\pi_i$ under the HALO2 ZK-SNARK scheme~\cite{halo2}. The proof attests that $\mathcal{M}_{\text{ref}}$ was executed correctly on $\mathcal{D}^{(i)}_{\text{test}}$ and that the claimed outputs $\mathbf{y}^{(i)}$ follow from that execution.

\vspace{0.5em}
\textit{Step 4: Proof Submission.} The validator submits the proof $\pi_i$ to the \texttt{Verifier SC}, together with the resulting output distribution, which the proof exposes as a public output.

\vspace{0.5em}
\textbf{Verification and Validator Admission.} After receiving $\pi_i$, the network initiates a structured verification procedure.
\texttt{Verifier SC} first validates the submitted proof using the embedded verification key. If verification is successful, the network computes the entropy-based quality score and the label-free class profile from the public output (Section~\ref{subsec:data_quality_assessment}). The candidate is admitted only if the proof is valid and the quality score exceeds the admission threshold.

\begin{algorithm*}[t]
\scriptsize
\caption{Validator Admission via Zero-Knowledge Attestation. The setting is fixed throughout: a single candidate validator $v$ runs the reference model $\mathcal{M}_{\text{ref}}$ on its private test set $\mathcal{D}_{\text{test}} = \{x_1,\dots,x_n\}$ of $n$ samples in the current round, and the model emits a probability vector over $C$ classes for each sample. We therefore drop the validator, model, and round indices and write all quantities for this candidate. The procedure either rejects $v$, or admits it together with a label-free class profile $\bar{\mathcal{P}}$.}
\label{alg:admission}
\begin{algorithmic}[1]
\Require candidate $v$ with private test set $\mathcal{D}_{\text{test}} = \{x_1,\dots,x_n\}$ \Comment{the $n$ samples $v$ never reveals}
\Require reference model $\mathcal{M}_{\text{ref}}$ over $C$ classes \Comment{same network model every candidate must run}
\Require proving key $pk$, on-chain verification key $vk$ \Comment{from the one-time circuit setup}
\Require entropy-quality threshold $\tau_Q$ \Comment{network-wide admission bar}
\Ensure reject $v$, or admit $v$ with class profile $\bar{\mathcal{P}}$
\State $\mathbf{y} \gets \big(\mathcal{M}_{\text{ref}}(x_1),\dots,\mathcal{M}_{\text{ref}}(x_n)\big)$ \Comment{run the model locally so raw data stays private}
\State commit to inputs, model, and outputs via Poseidon/KZG \Comment{bind $v$ to exactly these values, unforgeably}
\State $\pi \gets \textsc{Prove}(pk,\,\mathcal{M}_{\text{ref}},\,\mathcal{D}_{\text{test}},\,\mathbf{y})$ \Comment{HALO2 proof that $\mathbf{y}$ came from $\mathcal{M}_{\text{ref}}$ on $\mathcal{D}_{\text{test}}$}
\State submit $\pi$ and the commitments to the \texttt{Verifier SC} \Comment{hand the proof to the on-chain checker}
\If{$\neg\,\textsc{Verify}(vk,\pi)$}
    \State \Return reject \Comment{execution not attested on-chain}
\EndIf
\State $Q \gets 1 - \dfrac{1}{n}\displaystyle\sum_{k=1}^{n}\dfrac{H\!\big(\mathrm{softmax}(\mathbf{y}_k)\big)}{\log C}$ \Comment{entropy-quality score in $[0,1]$, Eq.~\eqref{eq:dqs-Q}}
\If{$Q < \tau_Q$}
    \State \Return reject \Comment{data not genuine: noise, out-of-distribution, or fabricated}
\EndIf
\State $\bar{\mathcal{P}} \gets \dfrac{1}{n}\sum_{k=1}^{n}\mathrm{softmax}\!\big(\mathbf{y}_k\big)$ \Comment{mean softmax over the $C$ classes gives the class makeup, no labels, Eq.~\eqref{eq:dqs-profile}}
\State \Return admit $v$ into $\mathcal{V}$ with profile $\bar{\mathcal{P}}$ \Comment{drives IID pool composition, Sec.~\ref{subsec:dqs_results}}
\end{algorithmic}
\end{algorithm*}

Algorithm~\ref{alg:admission} summarizes the complete admission procedure. Admitted validators therefore provide both a verifiable computation trace and a privacy-preserving estimate of dataset quality, which serves as the basis for the masking and consensus mechanisms described next.

\subsubsection{Validator Data Quality Assessment}
\label{subsec:data_quality_assessment}

To assess a validator's dataset without exposing its samples, we analyze the probability distributions produced by the reference model rather than hard predictions. Since accuracy requires ground-truth labels that remain private, we instead derive two quantities from the output distributions: an entropy-based quality score for admission and a label-free class profile. Consider a validator $j \in V$ that evaluates a model $i$ on a private test set $TS_j$. For each data point $d \in TS_j$, the model produces a probability distribution:
\begin{equation}
    \mathcal{P}_i^d = [p_1^d, p_2^d, \dots, p_n^d], \quad \text{s.t.}\quad \sum_{c=1}^{n} p_c^d = 1.
\end{equation}
\noindent
We first quantify prediction uncertainty through Shannon entropy:
\begin{equation}
    H(\mathcal{P}_i^d) = -\sum_{c=1}^{n} p_c^d \log(p_c^d),
\end{equation}
where lower entropy values imply higher certainty and stronger dataset representativeness. The entropy-based data quality score $Q_{ij}$ for validator $j$ is defined as:
\begin{equation}
    \label{eq:dqs-Q}
    Q_{ij} = 1 - \frac{1}{|TS_j|}\sum_{d \in TS_j}\frac{H(\mathcal{P}_i^d)}{\log(n)},
\end{equation}
The score is normalized to the interval $[0,1]$. A candidate is admitted only if $Q_{ij}\ge\tau_Q$, filtering out noisy, fabricated, or out-of-distribution data that typically produce high-entropy predictions. This criterion corresponds to the performance assessment of Section~\ref{subsec:validator_selection_zkp}. The class imbalance alone does not reduce $Q_{ij}$, as confident predictions on a subset of classes remain informative.

The mean output distribution defines the validator's class profile:
\begin{equation}
    \label{eq:dqs-profile}
    \bar{\mathcal{P}}_{ij} = \frac{1}{|TS_j|}\sum_{d \in TS_j}\mathcal{P}_i^d,
\end{equation}
which estimates the class composition of $TS_j$ without revealing labels. Together with the admission proof, this profile enables the network to assemble a validator set whose aggregate class distribution is close to IID while never accessing raw data.

Both the entropy score and the class profile are computed solely from the probability distribution of the output, preserving data privacy. The zero-knowledge proof certifies correct model execution on the committed inputs. The entropy score provides a confidence-based proxy for dataset suitability, while the mean output distribution estimates its class composition. These signals do not certify sample authenticity, label correctness, or representativeness..

\subsection{Gradient-Based Mask Generation}
\label{subsec:gradient_mask_generation}

Only admitted validators perform the gradient-based masking strategy of Arazzi et al.~\cite{arazzi2025privacy}. During each training round, the validators evaluate the submitted client models on their private validation data and compute model-specific masks. In each training round $t$, validators receive multiple locally trained models submitted by the clients to a shared model pool. Validators independently evaluate each submitted model on their local validation datasets and compute model-specific masks from the results.\vspace{0.5em}

\textbf{Class-wise reliability scoring.} We consider a validator $v \in \mathcal{V}$ receiving a set of candidate models $\mathcal{M}^{(t)} = \{\mathbf{M}_1^{(t)}, \mathbf{M}_2^{(t)}, \dots, \mathbf{M}_K^{(t)}\}$ at round $t$. For each candidate $\mathbf{M}_k^{(t)}$, the validator $v$ first measures the class-wise predictive reliability on its private validation set:
\begin{equation}
A^{(t)}_{k,v,c} = \text{Accuracy}\left(\mathbf{M}_k^{(t)}, \mathcal{D}^{\text{val}}_{v,c}\right),
\end{equation}
where $\mathcal{D}^{\text{val}}_{v,c}$ denotes the subset of validator $v$'s validation data in class $c$. The mask is generated from a reliability-weighted loss over the full validation set, which weights each sample by the model's measured reliability on the class of that sample:
\begin{equation}
\label{eq:mask-loss}
\mathcal{L}^{\text{mask}}_{k,v} = \frac{1}{|\mathcal{D}^{\text{val}}_{v}|} \sum_{(\mathbf{x},y) \in \mathcal{D}^{\text{val}}_{v}} A^{(t)}_{k,v,y}\, \ell\left(f(\mathbf{x}; \mathbf{M}_k^{(t)}), y\right).
\end{equation}
Class-wise weighting keeps the mask informative across heterogeneous data.\vspace{0.5em}

\textbf{Saliency and adaptive retention.} The saliency of each trainable coordinate is its absolute gradient under this loss:
\begin{equation}
\mathbf{G}^{(t)}_{k,v} = \left| \nabla_{\mathbf{M}_k^{(t)}} \mathcal{L}^{\text{mask}}_{k,v} \right|.
\end{equation}
Let $\mathbf{G}_{(1)} \geq \mathbf{G}_{(2)} \geq \dots \geq \mathbf{G}_{(P)}$ be the ordered gradient magnitudes of $P$. Retention is adaptive: the mask keeps the fewest coordinates that together capture a fraction $\eta$ of the total gradient energy,
\begin{equation}
k_{\eta} = \min\left\{ m : \sum_{j=1}^{m} \mathbf{G}_{(j)} \geq \eta \sum_{j=1}^{P} \mathbf{G}_{(j)} \right\}.
\end{equation}
To avoid degenerate masks, the keep count is clamped between a minimum and a maximum retention ratio:
\begin{equation}
k = \min\left( \lfloor r_{\max} P \rfloor,\ \max\left( \lceil r_{\min} P \rceil,\ k_{\eta} \right) \right).
\end{equation}
The retention threshold $\tau^{(t)}_{k,v} = \mathbf{G}_{(k)}$ is the smallest magnitude among the retained coordinates. The default configuration uses $\eta=0.9$, $r_{\min}=0.05$, and $r_{\max}=0.8$. The energy threshold preserves the most informative gradients, while the retention bounds prevent excessively sparse or dense masks.

\textbf{Adaptive attenuation.} The mask preserves the retained coordinates at full strength and attenuates the rest:
\begin{equation}
\mathbf{Mask}^{(t)}_{k,v}[i] = \begin{cases}
1 & \text{if } \mathbf{G}^{(t)}_{k,v}[i] \geq \tau^{(t)}_{k,v},\\
\gamma^{(t)}_{k,v} & \text{otherwise}.
\end{cases}
\end{equation}
The scale-down factor $\gamma^{(t)}_{k,v}$ is also read from the gradient distribution as the ratio of the median discarded magnitude to the median retained magnitude:
\begin{equation}
\gamma^{(t)}_{k,v} = \min\left( 1,\ \frac{\text{median}(\mathbf{G}_{\text{discard}})}{\max\left(\text{median}(\mathbf{G}_{\text{keep}}), \epsilon\right)} \right),
\end{equation}
This adaptive attenuation reflects the separation between retained and discarded gradients while remaining robust to heavy-tailed distributions.

\textbf{Optional smoothing.} Masks may be exponentially smoothed across rounds for the same client-validator pair:
\begin{equation}
\mathbf{Mask}^{(t)}_{k,v} \leftarrow (1 - \beta) \mathbf{Mask}^{(t)}_{k,v} + \beta \mathbf{Mask}^{(t-1)}_{k,v},
\end{equation}
where $\beta \in [0,1]$ trades responsiveness against stability. where $\beta\in[0,1]$ controls temporal smoothing. Our experiments use $\beta=0$.\vspace{0.5em}

\textbf{Mask application.} Finally, the mask scales the client's update direction rather than the raw weights, so retained coordinates pass unchanged and the rest are attenuated toward the current global model $\mathbf{w}^{(t)}$:
\begin{equation}
\tilde{\mathbf{M}}^{(t)}_{k,v} = \mathbf{w}^{(t)} + \mathbf{Mask}^{(t)}_{k,v} \odot \left( \mathbf{M}^{(t)}_{k} - \mathbf{w}^{(t)} \right),
\end{equation}
where $\odot$ denotes element-wise multiplication. The resulting masked updates are used during decentralized aggregation (Section~\ref{subsec:bft_consensus}). The complete routine is given in Algorithm~\ref{alg:masking}.

\begin{algorithm*}[t]
\scriptsize
\caption{Gradient-Based Mask Generation for one pooled model. All quantities are computed by a single fixed validator $v$ for one candidate model at one round $t$ (written $\mathbf{M}^{(t)}_k$ elsewhere), so the validator $v$, model index $k$, and round superscript $(t)$ are dropped from the symbols inside the body. Here $\mathbf{M}$ is the pooled client model, $\mathbf{w}$ is the current global model, and $P$ is the number of trainable coordinates.}
\label{alg:masking}
\begin{algorithmic}[1]
\Require pooled model $\mathbf{M}$; global model $\mathbf{w}$
\Require validator's private validation set $\mathcal{D}^{\text{val}}$, with per-class subsets $\mathcal{D}^{\text{val}}_{c}$
\Require energy fraction $\eta$ \Comment{share of gradient energy the mask must keep}
\Require retention clamps $r_{\min}, r_{\max}$ \Comment{floor and ceiling on the kept fraction of $P$}
\Require smoothing weight $\beta$ \Comment{$\beta = 0$ disables cross-round smoothing}
\Require previous-round mask $\mathbf{Mask}_{\text{prev}}$ \Comment{used only when $\beta > 0$}
\Require numerical floor $\epsilon$
\Ensure masked update $\tilde{\mathbf{M}}$
\For{each class $c$}
    \State $A_{c} \gets \textsc{Accuracy}(\mathbf{M},\,\mathcal{D}^{\text{val}}_{c})$ \Comment{how reliable $\mathbf{M}$ is on class $c$}
\EndFor
\State $\mathcal{L} \gets \dfrac{1}{|\mathcal{D}^{\text{val}}|}\displaystyle\sum_{(\mathbf{x},y)\,\in\,\mathcal{D}^{\text{val}}} A_{y}\,\ell\!\big(f(\mathbf{x};\mathbf{M}),\,y\big)$ \Comment{reliability-weighted loss, Eq.~\eqref{eq:mask-loss}}
\State $\mathbf{G} \gets \big|\nabla_{\mathbf{M}}\,\mathcal{L}\big|$ \Comment{per-coordinate saliency is the absolute gradient}
\State sort the magnitudes so that $\mathbf{G}_{(1)} \ge \mathbf{G}_{(2)} \ge \cdots \ge \mathbf{G}_{(P)}$ \Comment{rank coordinates from most to least salient}
\State $k_{\eta} \gets \min\Big\{\, m \;:\; \textstyle\sum_{j=1}^{m}\mathbf{G}_{(j)} \,\ge\, \eta\sum_{j=1}^{P}\mathbf{G}_{(j)} \,\Big\}$ \Comment{fewest top coordinates holding fraction $\eta$ of the energy}
\State $K_{\text{keep}} \gets \min\!\big(\lfloor r_{\max}P\rfloor,\ \max(\lceil r_{\min}P\rceil,\ k_{\eta})\big)$ \Comment{clamp keep count to $[\lceil r_{\min}P\rceil,\ \lfloor r_{\max}P\rfloor]$}
\State $\tau \gets \mathbf{G}_{(K_{\text{keep}})}$ \Comment{threshold is the smallest kept magnitude}
\State $\mathbf{G}_{\text{keep}} \gets \{\,\mathbf{G}_{(j)} : j \le K_{\text{keep}}\,\}$ \Comment{the retained magnitudes}
\State $\mathbf{G}_{\text{discard}} \gets \{\,\mathbf{G}_{(j)} : j > K_{\text{keep}}\,\}$ \Comment{the remaining magnitudes}
\State $\gamma \gets \min\!\Big(1,\ \dfrac{\mathrm{median}(\mathbf{G}_{\text{discard}})}{\max\!\big(\mathrm{median}(\mathbf{G}_{\text{keep}}),\,\epsilon\big)}\Big)$ \Comment{how far discarded coordinates sit below retained ones}
\For{each coordinate $i \in \{1,\dots,P\}$}
    \If{$\mathbf{G}[i] \ge \tau$}
        \State $\mathbf{Mask}[i] \gets 1$ \Comment{keep this coordinate at full strength}
    \Else
        \State $\mathbf{Mask}[i] \gets \gamma$ \Comment{attenuate this coordinate}
    \EndIf
\EndFor
\If{$\beta > 0$}
    \State $\mathbf{Mask} \gets (1-\beta)\,\mathbf{Mask} + \beta\,\mathbf{Mask}_{\text{prev}}$ \Comment{optional EMA smoothing across rounds}
\EndIf
\State \Return $\tilde{\mathbf{M}} \gets \mathbf{w} + \mathbf{Mask}\odot(\mathbf{M} - \mathbf{w})$ \Comment{scale the update direction $\mathbf{M}-\mathbf{w}$, not the raw weights; $\odot$ is element-wise}
\end{algorithmic}
\end{algorithm*}

\subsection{Challenge-based Mask Verification}
\label{subsec:challenge_verification}

\begin{figure*}[t]
    \centering
    \includegraphics[width=\linewidth]{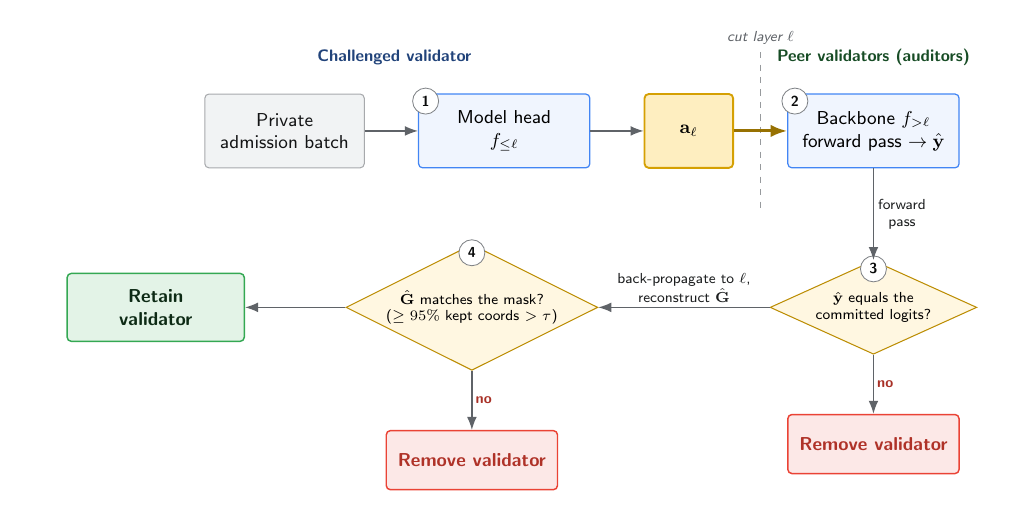}
    \caption{Challenge-based mask verification as an interaction between the challenged validator and the peer auditors, split by the cut layer~$\ell$. The validator re-runs the model head $f_{\le\ell}$ on its private admission batch and exposes only the smashed activations $\mathbf{a}_\ell$, the single quantity that crosses the cut layer. Peers complete the forward pass and remove the validator if the recomputed logits disagree with its committed logits; otherwise they back-propagate to layer~$\ell$, reconstruct the gradient $\hat{\mathbf{G}}$, and retain the validator only when at least 95\% of its kept mask coordinates stay above the published threshold~$\tau$ with none scaled down.}
    \label{fig:challenge_verification}
\end{figure*}

Zero-knowledge admission attests to the validator's dataset at admission time, but cannot prevent later manipulation during mask generation. To preserve privacy while still detecting this behavior, we embed a cut-layer audit inspired by Split Learning~\cite{vepakomma2018split} and illustrated in Figure~\ref{fig:challenge_verification}. At scheduled checkpoints, or whenever on-chain heuristics flag suspicious updates, the network requests a validator to re-run the \emph{same} batch that was used in its admission proof and to expose only the intermediate activations at a pre-defined cut layer (the smashed data). Other validators feed these activations through the remaining layers of the public model. They confirm that the logits match those already committed under the validator's most recent ZKP, then back-propagate to the cut layer to obtain an independent estimate of the gradient tensor. This reconstructed gradient is compared with the validator's published mask, which must include two public items:  

\[
\tau^{(t)}_{k,v}
  = \operatorname{top\text{-}k}\!\bigl(|\mathbf{G}^{(t)}_{k,v}|\bigr),\quad
  k=\lfloor p\cdot|\mathbf{G}^{(t)}_{k,v}|\rfloor,
\]
and the unsigned magnitudes of every parameter that the mask keeps at full scale. The validator publishes the pruning threshold $\tau^{(t)}_{k,v}$ together with the unsigned magnitudes of all the parameters retained. A mask is accepted if at least 95\% of the retained parameters remain above $\tau^{(t)}_{k,v}$ in the reconstructed gradient and no parameter above the threshold has been attenuated. Because only smashed activations and retained gradient magnitudes are disclosed, raw data remain private while manipulated masks become detectable. Algorithm~\ref{alg:challenge} states the audit.

\begin{algorithm*}[t]
\scriptsize
\caption{Challenge-Based Mask Verification. One challenged validator $v$ is audited at training round $t$ for one pooled model with index $k$. The indices $v$, $t$, and $k$ are fixed throughout, so the body writes the threshold as $\tau$, the reconstructed gradient as $\hat{\mathbf{G}}$, and the peer logits as $\hat{\mathbf{y}}$ without subscripts. The audit reconstructs the gradient at the cut layer from a privacy-preserving replay and keeps $v$ only if its published mask matches the peer-reconstructed gradient.}
\label{alg:challenge}
\begin{algorithmic}[1]
\Require Challenged validator $v$ \Comment{the node whose mask is under suspicion}
\Require Cut layer $\ell$ \Comment{boundary splitting the public model into the model head $f_{\le \ell}$ and the backbone $f_{>\ell}$}
\Require Published threshold $\tau$ and retained magnitudes \Comment{top-$p\%$ cutoff, Sec.~\ref{subsec:gradient_mask_generation}}
\Require Agreement bound $\theta = 0.95$ \Comment{minimum fraction of retained coordinates that must be justified}
\Ensure Decision to retain or remove $v$
\State $v$ re-runs its admission batch on $f_{\le \ell}$ \Comment{same batch attested in the ZKP}
\State $v$ exposes only the cut-layer activations $\mathbf{a}_\ell$ \Comment{smashed data, so the raw inputs stay private}
\State peers complete the forward pass $\hat{\mathbf{y}} \gets f_{>\ell}(\mathbf{a}_\ell)$ \Comment{independently recompute the model outputs}
\If{$\hat{\mathbf{y}}$ disagrees with the committed logits}
    \State \Return remove $v$ \Comment{replay is inconsistent with the ZKP}
\EndIf
\State peers back-propagate to layer $\ell$ to reconstruct the gradient $\hat{\mathbf{G}}$ \Comment{independent estimate of the true gradient}
\State $r \gets$ fraction of retained coordinates $i$ with $\hat{\mathbf{G}}[i] \ge \tau$ \Comment{how much of the kept mask the gradient supports}
\If{$r \ge \theta$ \textbf{and} no coordinate above $\tau$ was scaled down}
    \State \Return retain $v$ \Comment{mask matches the reconstructed gradient}
\Else
    \State \Return remove $v$ \Comment{mask not justified by the reconstructed gradient}
\EndIf
\end{algorithmic}
\end{algorithm*}

\subsection{Byzantine Fault Tolerant Consensus Implementation}
\label{subsec:bft_consensus}
\label{sec:mask-consensus}

For each submitted client model $LM_k$, every validator $v_i \in \mathcal{V}$ independently generates a candidate mask $M_{k,i}$. Rather than averaging these masks, the protocol selects a single canonical mask through quality-aware BFT consensus. The protocol consists of four steps.

\vspace{0.5em}
\noindent\textbf{Step 1: Local Evaluation of All Candidate Masks.} Each validator $v_i$ evaluates every candidate mask $M_{k,j}$ submitted for model $LM_k$ by applying the mask to the model and computing a performance metric $s_{k,j}^{(i)}$ using its own held-out test set $\mathcal{D}_{\text{val}}^{(i)}$:
\begin{equation}
    s_{k,j}^{(i)} = \text{Performance}(LM_k \odot M_{k,j}, \mathcal{D}_{\text{val}}^{(i)}),
\end{equation}
where $\odot$ denotes the element-wise application of the mask to the model parameters. The performance function can represent classification accuracy, masked loss, F1-score, or other task-specific objectives. To formalize the evaluation structure, we define the validator score matrix $\mathbf{S}_k \in \mathbb{R}^{N \times N}$ for each model $LM_k$ as:
\begin{equation}
    \mathbf{S}_k = \left[s_{k,j}^{(i)}\right]_{i,j=1}^{N},
\end{equation}
where the $i$-th row corresponds to scores computed by the validator $v_i$ and the $j$-th column aggregates the scores of mask $M_{k,j}$ across validators.

\vspace{0.5em}
\noindent\textbf{Step 2: Score Aggregation.} Each validator locally computes the average score $\bar{s}_{k,j}$ for each mask $M_{k,j}$ as:
\begin{equation}
    \bar{s}_{k,j} = \frac{1}{N} \sum_{i=1}^N s_{k,j}^{(i)}.
\end{equation}
A median score is a drop-in alternative that further dampens scoring manipulation by adversarial validators, at the cost of discarding magnitude information:
\begin{equation}
    \bar{s}_{k,j} = \text{median}_{i \in \{1, \dots, N\}} \left(s_{k,j}^{(i)}\right),
\end{equation}

\vspace{0.5em}
\noindent\textbf{Step 3: Local Selection of Optimal Mask.} Validators independently determine the most promising mask for model $LM_k$ using:
\begin{equation}
    M_k^\star = \arg\max_{j \in \{1,\dots,N\}} \bar{s}_{k,j}.
\end{equation}
Computing \mbox{$M_k^\star$} locally at each validator, rather than trusting a single coordinator, removes any single point of control; BFT finalization in Step~4 then limits the influence of up to \mbox{$f$} dishonest validators.

\vspace{0.5em}
\noindent\textbf{Step 4: BFT Finalization.} The selected mask $M_k^\star$ is proposed in a BFT consensus instance. A designated proposer submits the candidate, and all validators verify that it matches their locally computed $M_k^\star$. Upon receiving $\geq 2f+1$ affirmative votes (assuming a validator set of $3f+1$), the mask is finalized and committed as the canonical mask for $LM_k$.

\vspace{0.5em}
\noindent\textbf{Global Model Aggregation.} Once a canonical mask $M_k^\star$ is committed for every model $LM_k$, the next global model is formed as a weighted average of the masked client models. Each winning mask scales its client's update direction as in Section~\ref{subsec:gradient_mask_generation}, $\tilde{\mathbf{M}}_k^\star = \mathbf{w}^{(t)} + M_k^\star \odot (LM_k - \mathbf{w}^{(t)})$. The standard FedAvg baseline weights each client by its local dataset size:
\begin{equation}
\label{eq:fedavg-weights}
\omega_k^{\text{FedAvg}} = \frac{|\mathcal{D}^{(k)}|}{\sum_{j=1}^{K} |\mathcal{D}^{(j)}|}.
\end{equation}
In contrast, our framework derives the aggregation weight of each client directly from the consensus evaluation, requiring no disclosure of the sizes of private dataset. Let $\bar{s}_k^\star = \max_{j} \bar{s}_{k,j}$ denote the score of the winning mask for model \mbox{$LM_k$}, that is, its validator-averaged score from Step~2. The consensus-driven weight is then
\begin{equation}
\label{eq:consensus-weights}
\omega_k^{\text{cons}} = \frac{\bar{s}_k^\star}{\sum_{j=1}^{K} \bar{s}_j^\star},
\end{equation}
so that the finalized global model is $\mathbf{w}^{(t+1)} = \sum_{k=1}^{K} \omega_k^{\text{cons}}\, \tilde{\mathbf{M}}_k^\star$. This score-based weighting implicitly down-weights low-quality or adversarial submissions without relying on client-reported metadata; its empirical behavior is analyzed in Section~\ref{sec:weight-evolution}. Algorithm~\ref{alg:consensus} summarizes consensus selection and aggregation.

\begin{algorithm*}[t]
\scriptsize
\caption{Quality-Aware BFT Mask Consensus and Global Aggregation.
For a fixed round $t$, every validator scores every candidate mask on its own held-out validation set, the per-mask scores are averaged across validators, the highest-scoring mask is selected and BFT-finalized per model, and each client's aggregation weight is set from its winning mask score rather than its dataset size.
Indices: $k$ ranges over the $K$ submitted client models, $i$ over the $N=3f+1$ validators, and $j$ over the candidate masks of the current model.}
\label{alg:consensus}
\begin{algorithmic}[1]
\Require submitted models $\{LM_k\}_{k=1}^{K}$
\Require validators $\mathcal{V}=\{v_1,\dots,v_N\}$ with $N=3f+1$, held-out val sets $\{\mathcal{D}^{(i)}_{\text{val}}\}$
\Require candidate masks $\{M_{k,j}\}$, one per validator per model, so $j=1,\dots,N$
\Require current global model $\mathbf{w}^{(t)}$
\Ensure next global model $\mathbf{w}^{(t+1)}$
\For{each model $LM_k$}
    \State \textbf{Reliability.} \Comment{score every candidate mask, then pick the best}
    \For{each validator $v_i$ and candidate mask $M_{k,j}$}
        \State $s^{(i)}_{k,j} \gets \textsc{Performance}\!\big(LM_k \odot M_{k,j},\,\mathcal{D}^{(i)}_{\text{val}}\big)$ \Comment{cross-validator scoring on private val data}
    \EndFor
    \For{each candidate mask $M_{k,j}$}
        \State $\bar{s}_{k,j} \gets \dfrac{1}{N}\sum_{i=1}^{N} s^{(i)}_{k,j}$ \Comment{mean over validators; median is a drop-in alternative}
    \EndFor
    \State $M^\star_k \gets \arg\max_{j} \bar{s}_{k,j}$ \Comment{each validator selects the best mask locally}
    \State $\bar{s}^\star_k \gets \max_{j} \bar{s}_{k,j}$ \Comment{mean score of the winning mask, reused as the weight}
    \State \textbf{Finalization.} \Comment{commit only with a Byzantine fault tolerant quorum}
    \If{$M^\star_k$ secures $\ge 2f+1$ matching votes out of $3f+1$}
        \State finalize $M^\star_k$ as the canonical mask for $LM_k$ \Comment{BFT commit}
    \EndIf
\EndFor
\State \textbf{Weighting.} \Comment{set client weights from mask quality, not dataset size}
\For{each model $LM_k$}
    \State $\omega_k \gets \dfrac{\bar{s}^\star_k}{\sum_{k'=1}^{K} \bar{s}^\star_{k'}}$ \Comment{consensus weight over models $k'$, Eq.~\eqref{eq:consensus-weights} vs.\ FedAvg Eq.~\eqref{eq:fedavg-weights}}
    \State $\tilde{\mathbf{M}}^\star_k \gets \mathbf{w}^{(t)} + M^\star_k \odot (LM_k - \mathbf{w}^{(t)})$ \Comment{masked update direction, Sec.~\ref{subsec:gradient_mask_generation}}
\EndFor
\State \textbf{Aggregation.} \Comment{quality-weighted global model}
\State $\mathbf{w}^{(t+1)} \gets \sum_{k=1}^{K} \omega_k\,\tilde{\mathbf{M}}^\star_k$
\State \Return $\mathbf{w}^{(t+1)}$
\end{algorithmic}
\end{algorithm*}

\vspace{0.5em}
\noindent\textbf{Security and Robustness Considerations.} Selecting masks by cross-validator score, and committing them under BFT, gives the protocol three properties. First, the winning mask is the one that most improves accuracy on the validators' private data, so selection rewards measured performance rather than a self-reported claim. Second, the BFT layer commits a mask only when at least \mbox{$2f+1$} of \mbox{$3f+1$} validators agree, which limits the influence of up to \mbox{$f$} Byzantine validators on the outcome. Third, because every score is computed locally on private validation data, the protocol stays decentralized and never exposes a validator's dataset. Together, performance-based selection and BFT finalization commit a high-quality mask for each client model while preserving privacy and resilience under the fault bound.

\subsection{Complexity Analysis}
\label{subsec:complexity}

We analyze the protocol asymptotically in terms of computation, communication, and storage. Let $N$ denote the number of clients, $V=3f+1$ the validator committee, $M$ the number of candidate models per round, $P\le V$ the number of validators participating in a challenge, $d$ the model dimension, $n$ the size of the validator test-set, $b$ the admission/challenge batch size, $F$ the cost of one forward pass (with backward pass $\Theta(F)$), and $S=\Theta(bF)$ the HALO2 circuit size. Table~\ref{tab:complexity} summarizes the results.

\textbf{Admission attestation, once per validator.} A candidate runs the reference model, $O(bF)$, commits its inputs and outputs with Poseidon and KZG, $O(b)$, and produces a HALO2 proof. For the system we use, the proof is quasi-linear in the circuit size, $\tilde{O}(S)=\tilde{O}(bF)$ time, and $O(S)$ space, while the proof and its verification are succinct, constant in $S$~\cite{halo2,kate2010constant}. Scoring the outputs adds $O(b)$. Overall, admission costs $\tilde{O}(bF)$ time and $O(bF)$ space, incurred once per validator and amortized over subsequent rounds.

\textbf{Evaluation and mask generation, per validator per round.} A validator evaluates each of the $M$ pooled models on its test set with one forward and one backward pass, $O(MnF)$, and selects the high-saliency coordinates of each mask by linear-time selection, $O(Md)$. The per-validator round cost is $O(M(nF+d))$, linear in both the pool size and the model dimension.

\textbf{Challenge audit, per challenge.} The challenged validator runs the model head to the cut layer, $O(bF)$, and exposes $O(b\lvert\mathcal{A}_\ell\rvert)$ bytes of activations, where $\lvert\mathcal{A}_\ell\rvert$ is the activation size at cut layer $\ell$. Each of the $P$ peers completes the forward pass, back-propagates the post-cut sub-model, and compares the reconstructed gradient against the published mask, $O(P(bF+d))$. The audit avoids a second zero-knowledge proof over the backward pass, eliminating an additional $\tilde{O}(bF)$ proving cost.

\textbf{Consensus, per round.} Consensus requires an all-to-all pre-vote/pre-commit exchange among the $V$ validators, giving $O(V^2)$ messages over a constant number of communication rounds. Combining $V$ masks into the consensus aggregation weights costs $O(Vd)$.

Overall, the recurring per-round cost is linear in pool size $M$ and model dimension $d$, and quadratic only in the validator committee size $V$. The only superlinear component is the one-time admission proof, whose $\tilde{O}(bF)$ time and $O(bF)$ space costs are amortized across all subsequent training rounds. The space overhead is discussed further in Section~\ref{sec:limitations}.

\begin{table}[t]
\centering
\caption{Asymptotic cost of the validator protocol. Admission is paid once per validator; evaluation, masking, and consensus are per round, and the audit is per challenge. Here $V=3f+1$ validators, $M$ pooled models, $P$ peers, $d$ parameters, $n$ test samples, $b$ batch, $F$ forward-pass cost, $S=\Theta(bF)$ circuit size, and $\lvert\mathcal{A}_\ell\rvert$ the cut-layer activation size.}
\label{tab:complexity}
\resizebox{\columnwidth}{!}{\begin{tabular}{llll}
\toprule
Phase & Time & Space & Communication \\
\midrule
Admission & $\tilde{O}(bF)$ & $O(bF)$ & $O(1)$ \\
Eval + masking & $O(M(nF+d))$ & $O(d)$ & $O(Md)$ \\
Challenge audit & $O(P(bF+d))$ & $O(d)$ & $O(b\lvert\mathcal{A}_\ell\rvert)$ \\
Consensus & $O(Vd)$ & $O(Vd)$ & $O(V^2)$ \\
\bottomrule
\end{tabular}}
\end{table}

\section{Experimental Setup}
\label{sec:setup}
We describe the configuration used throughout the evaluation: the datasets and their partitioning, the model and training protocol, the attack we study, the evaluation metrics, the FedAvg baseline and the zero-knowledge attestation pipeline.

\subsection{Datasets}
We evaluate our framework on four widely adopted datasets: CIFAR-10~\cite{krizhevsky2009learning}, CIFAR-100~\cite{krizhevsky2009learning}, FashionMNIST~\cite{xiao2017fashion}, and GTSRB~\cite{stallkamp2012man}. These datasets were selected due to their extensive use in federated and decentralized learning research~\cite{xie2019dba,walter2024mitigating,li2021decentralized,reyes2025precision,konevcny2016strategies}, enabling direct comparability with established baselines. CIFAR-10 comprises 60,000 32$\times$32 color images evenly distributed across 10 object classes, with 50,000 samples allocated for training and 10,000 for testing. CIFAR-100 follows a similar structure, but introduces greater task complexity by expanding the class set to 100 categories. FashionMNIST consists of 70,000 grayscale images (28$\times$28 pixels) spanning 10 categories of clothing items, split into 60,000 training and 10,000 test instances. GTSRB (the German Traffic Sign Recognition Benchmark) contains roughly 50,000 cropped color images of traffic signs across 43 classes, captured under varied real-world lighting and viewpoint conditions; its naturally imbalanced, fine-grained class structure complements the balanced object- and clothing-recognition tasks. To emulate real-world data heterogeneity, we partition each dataset across clients using a Dirichlet distribution, consistent with previous works~\cite{li2022federated}. We consider multiple values of the Dirichlet concentration parameter, $\alpha \in \{0.1, 0.3, 1.0\}$, together with an IID partition ($\alpha\!\to\!\infty$); these control the degree of non-IID-ness in the client data distributions. Smaller $\alpha$ values produce a higher per-client class imbalance, stressing the masking and validation mechanisms under non-uniform data.

\subsection{Model Architecture and Training}
In all experiments, we employ ResNet-18 as the base architecture, consistent with standard practices in image classification tasks under decentralized learning. The model is initialized with random weights and trained independently by each client using Stochastic Gradient Descent (SGD). The optimizer is configured with a learning rate of 0.05 and a momentum of 0.9, without weight decay. Local training is carried out with a batch size of 64 over a single local epoch per round, and the Swarm Learning process runs for 100 global rounds unless otherwise specified (the ablation studies in Section~\ref{sec:ablation} extend training to 250 rounds). The framework is implemented in PyTorch, and every experiment runs on a single NVIDIA GeForce RTX 4090 GPU (24~GB) with CUDA 13.2. This setup keeps the compute budget within a single GPU while covering the heterogeneity levels and topologies that the evaluation requires.

\subsection{Attack Model}
\label{sub:attack_models}
Adversarial behavior is applied only to training clients; validators are not poisoned in these experiments. The attack we evaluate is strictly \emph{client-local}: each malicious client operates solely on its own model update, with no knowledge of other clients' submissions or of the global model's history beyond the last published checkpoint. This constraint reflects the decentralized setting, where peer updates are not observable before aggregation.

We focus on \textbf{gradient scaling}~\cite{fang2020local}, the model-poisoning attack most directly aligned with the masking mechanism. A malicious client multiplies its update delta $\Delta w = w_{\text{local}} - w_{\text{global}}$ by a scale factor (10$\times$ by default) before submission, amplifying its influence on the aggregated model without altering the update direction. This stresses the saliency-based mask directly: it tests whether evaluation on validator data attenuates the disproportionately inflated coordinates that a magnitude-based average would otherwise propagate. Therefore, the robustness results in this paper are scoped to gradient scaling; we leave a broader suite of client-local attacks (label flipping, sign flipping, additive noise, and the same-value attack) to future work.

The default factor of 10$\times$ is not arbitrary. To find the scale at which the attack becomes effective, we sweep the factor over $\{2, 5, 10, 50\}$ on FashionMNIST with the rest of the configuration fixed (20\% attackers, $\alpha\!=\!0.3$), and measure the accuracy the undefended FedAvg baseline retains (Table~\ref{tab:scale_sweep}). The damage is a step change rather than a gradual decline. At 2$\times$ and 5$\times$ the baseline still retains at least 99.4\% of its clean accuracy, because a mildly inflated update survives averaging without dominating it. At 10$\times$ the same baseline collapses to chance, retaining only 11\%, and a larger 50$\times$ factor is partly self-defeating, since an update that extreme is averaged back toward irrelevance and retention recovers to 21\%. We therefore fix 10$\times$ as the default, the smallest scale at which gradient scaling reliably breaks an undefended aggregator, which is the regime in which a defense must prove itself.

\begin{table}[t]
\centering
\caption{Scale-factor sweep for the gradient-scaling attack on FashionMNIST (20\% attackers, $\alpha=0.3$). \emph{Accuracy retained} is the best accuracy under attack relative to the no-attack FedAvg baseline; higher means the attack did less damage. The undefended baseline tolerates small factors and collapses at 10$\times$, the value we adopt as the default.}
\label{tab:scale_sweep}
\begin{tabular}{lcccc}
\toprule
Scale factor & 2$\times$ & 5$\times$ & 10$\times$ & 50$\times$ \\
\midrule
Accuracy retained & 99.5\% & 99.4\% & 11\% & 21\% \\
\bottomrule
\end{tabular}
\end{table}

Unless otherwise noted, adversarial experiments fix the data heterogeneity at $\alpha\!=\!0.3$ and inject attackers at fractions of 10\%, 20\%, and 30\% of the training clients. Every adversarial configuration is run under both our mask-consensus protocol and the FedAvg baseline, isolating the contribution of validator scoring and consensus-driven aggregation.

\subsection{Evaluation Metrics}
Performance is measured using \emph{test accuracy} on clean data. Under attack we report \emph{accuracy retained}, defined as the best test accuracy achieved under attack normalized by the no-attack FedAvg baseline for the same topology (Section~\ref{sub:scaling_robustness}).

\subsection{Baseline}
The baseline is FedAvg under the same role-based data split. For this baseline, validator masking and consensus are disabled, and client submissions are averaged with dataset-size weights as in Eq.~\ref{eq:fedavg-weights}. This same-split comparison isolates the effect of validator scoring, mask selection, and consensus-derived aggregation weights.

\subsection{Zero-Knowledge Attestation Setup}
\label{subsec:zk_setup}
We benchmark the validator-admission attestation as a standalone pipeline. For each candidate dataset we train a compact reference classifier, export it to ONNX, and compile it into a HALO2 circuit with EZKL; the candidate then proves correct execution of this circuit on its private data, with Poseidon commitments binding the inputs, the model, and the outputs. Unless noted, the reference classifier is a single-block convolutional network (the \emph{SimpleCNN} probe); deeper ResNet-9 and ResNet-18 references exceed single-machine proving memory, as Section~\ref{sec:limitations} discusses. We measure cost and fidelity along four axes: proving and verification time, proof and circuit size, the accuracy lost to circuit quantization, and the overhead of the challenge-based mask audit. The quantization settings and circuit parameters are stated with each result.

\section{Experimental Results}
\label{sec:experiments}
We first establish clean-data performance against the FedAvg baseline, then examine the consensus-derived aggregation weights, ablate the two main design axes, quantify robustness to gradient-scaling poisoning, and finally report the cost of the zero-knowledge attestation and challenge protocol.

\subsection{Clean-Data Performance}
\label{sub:experiment_performance}

We first compare clean-data accuracy between our mask-consensus framework and the FedAvg baseline under identical role-based data splits. Table~\ref{tab:comparison} reports the best test accuracy reached over training for two configurations of our method, namely validators drawn from an IID pool with non-IID local clients (p-iid/l-niid) and the fully IID variant (p-iid/l-iid), across three network topologies and four datasets.

\begin{table*}[htbp]
\centering
\caption{FedAvg baseline vs.\ Mask Consensus (Ours). Best accuracy (\%). Ours variants: pool-iid/local-non\_iid, pool-iid/local-iid. Bold\,=\,winner per row-topology.}
\label{tab:comparison}
\small
\begin{tabular}{llccccccccc}
\toprule
 &  & \multicolumn{3}{c}{$N\!=\!10,V\!=\!3$} & \multicolumn{3}{c}{$N\!=\!20,V\!=\!5$} & \multicolumn{3}{c}{$N\!=\!30,V\!=\!6$} \\
\cmidrule(lr){3-5} \cmidrule(lr){6-8} \cmidrule(lr){9-11}
Dataset & $\alpha$ & FedAvg & \makecell{p-iid\\l-niid} & \makecell{p-iid\\l-iid} & FedAvg & \makecell{p-iid\\l-niid} & \makecell{p-iid\\l-iid} & FedAvg & \makecell{p-iid\\l-niid} & \makecell{p-iid\\l-iid} \\
\midrule
F-MNIST & 0.1 & $86.5$ & $85.3$ & $\mathbf{87.0}$ & $88.6$ & $\mathbf{89.0}$ & $88.5$ & $\mathbf{87.6}$ & $84.9$ & $85.4$ \\
 & 0.3 & $89.7$ & $\mathbf{89.8}$ & $89.4$ & $89.5$ & $90.2$ & $\mathbf{90.9}$ & $89.6$ & $89.7$ & $\mathbf{90.8}$ \\
 & 1.0 & $91.4$ & $\mathbf{92.0}$ & $91.8$ & $90.9$ & $\mathbf{91.3}$ & $91.0$ & $90.7$ & $91.1$ & $\mathbf{91.2}$ \\
 & IID & $91.8$ & NA & $\mathbf{92.0}$ & $91.6$ & NA & $\mathbf{92.0}$ & $91.5$ & NA & $\mathbf{91.5}$ \\
\midrule
CIFAR-10 & 0.1 & $63.7$ & $61.9$ & $\mathbf{64.0}$ & $63.7$ & $68.7$ & $\mathbf{76.1}$ & $41.4$ & $\mathbf{52.9}$ & $51.4$ \\
 & 0.3 & $78.7$ & $\mathbf{79.4}$ & $79.2$ & $\mathbf{78.6}$ & $76.9$ & $75.9$ & $\mathbf{78.9}$ & $67.9$ & $69.4$ \\
 & 1.0 & $77.8$ & $\mathbf{81.6}$ & $80.5$ & $76.5$ & $78.7$ & $\mathbf{79.4}$ & $\mathbf{79.5}$ & $76.2$ & $75.7$ \\
 & IID & $80.9$ & NA & $\mathbf{81.7}$ & $77.0$ & NA & $\mathbf{79.0}$ & $77.3$ & NA & $\mathbf{79.8}$ \\
\midrule
CIFAR-100 & 0.1 & $\mathbf{47.0}$ & $46.5$ & $45.6$ & $\mathbf{42.4}$ & $40.0$ & $40.4$ & $\mathbf{38.8}$ & $34.9$ & $35.0$ \\
 & 0.3 & $\mathbf{48.2}$ & $47.7$ & $48.0$ & $\mathbf{44.5}$ & $44.5$ & $44.2$ & $\mathbf{42.5}$ & $39.9$ & $40.0$ \\
 & 1.0 & $49.8$ & $\mathbf{50.0}$ & $49.8$ & $46.6$ & $45.7$ & $\mathbf{47.1}$ & $\mathbf{45.1}$ & $42.2$ & $42.5$ \\
 & IID & $\mathbf{50.8}$ & NA & $50.4$ & $\mathbf{48.2}$ & NA & $47.1$ & $44.8$ & NA & $\mathbf{45.0}$ \\
\midrule
GTSRB & 0.1 & $\mathbf{95.5}$ & $92.9$ & $92.6$ & $\mathbf{94.5}$ & $92.6$ & $92.0$ & $92.7$ & $92.4$ & $\mathbf{93.4}$ \\
 & 0.3 & $\mathbf{93.3}$ & $92.0$ & $91.9$ & $\mathbf{95.2}$ & $93.5$ & $93.9$ & $\mathbf{95.8}$ & $95.1$ & $94.6$ \\
 & 1.0 & $\mathbf{95.5}$ & $91.4$ & $92.8$ & $94.8$ & $\mathbf{95.1}$ & $93.9$ & $95.0$ & $96.0$ & $\mathbf{96.0}$ \\
 & IID & $\mathbf{93.5}$ & NA & $91.6$ & $\mathbf{94.6}$ & NA & $93.9$ & $93.9$ & NA & $\mathbf{96.2}$ \\
\bottomrule
\end{tabular}
\end{table*}

Two patterns are visible in Table~\ref{tab:comparison}. First, the advantage of mask consensus is largest under high data heterogeneity, and it widens with network scale: on CIFAR-10 at $\alpha\!=\!0.1$, accuracy rises from $63.7\%$ for FedAvg to $76.1\%$ at $N\!=\!20$, and from $41.4\%$ to $52.9\%$ at $N\!=\!30$, where FedAvg degrades sharply as each client sees a narrower slice of the label space. Second, on the easier, balanced FashionMNIST task both methods cluster within one or two points, confirming that the masking machinery neither helps nor harms when heterogeneity is mild.
On the balanced or already-high-accuracy datasets the two methods track each other closely, which is the intended behavior. GTSRB stays above \mbox{$90\%$} even under skew, so the saliency-based mask has little low-quality signal to filter and consensus weighting matches FedAvg rather than improving on it; on CIFAR-100 the two stay within one to four points across settings. Mask consensus is built to protect aggregation under heterogeneity and attack, not to raise clean accuracy where every client already trains a strong model, and the clean-data comparison confirms it adds that protection without sacrificing accuracy. Its clean-data gains concentrate on heterogeneous CIFAR-10, while its core benefits, robustness and verifiable admission, are demonstrated in the sections that follow.
Across the four benchmarks the framework holds clean-data accuracy on par with FedAvg, with a clear advantage on the hardest non-IID partitions (up to roughly 12 points on CIFAR-10 at \mbox{$\alpha\!=\!0.1$}) and a gap of at most a few points on CIFAR-100 and the already-saturated GTSRB. Its robustness and verifiable-admission guarantees therefore come at no measurable cost to clean accuracy.

\subsection{Consensus-Driven Aggregation Weights}
\label{sec:weight-evolution}

We examine how the consensus-derived aggregation weights behave in practice compared to the static dataset-size-proportional weights used in standard FedAvg.  Recall from Section~\ref{sec:mask-consensus} that each client's aggregation weight is determined by the mean validator score of its winning mask proposal (Eq.~\ref{eq:consensus-weights}), rather than by its local dataset size (Eq.~\ref{eq:fedavg-weights}).

\begin{figure}[t]
  \centering
  \includegraphics[width=\columnwidth]{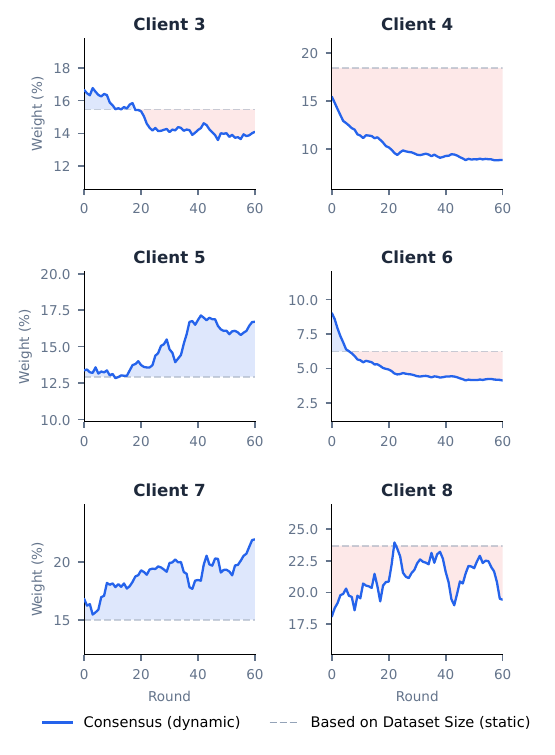}
  \caption{Per-client aggregation weight over 61 rounds (CIFAR-10, $\alpha\!=\!0.1$, $N\!=\!10$, $V\!=\!3$).  Solid blue: consensus-derived weight; dashed grey: static FedAvg weight proportional to local dataset size.  Blue shading marks rounds where a client receives \emph{more} influence than its data size would warrant; red shading marks the opposite.}
  \label{fig:weight-evolution}
\end{figure}

Figure~\ref{fig:weight-evolution} tracks these weights across 61 rounds for six of the seven training clients under a strongly non-IID setting (CIFAR-10, $\alpha=0.1$, $N\!=\!10$, $V\!=\!3$).  Two clear patterns emerge. The consensus weights depart from the static FedAvg baseline (dashed lines).  Client 8, which holds the largest partition and would receive 23.7\% under FedAvg, is progressively down-weighted to 17.4\% by round 60, as validators judge its submissions as less useful than its data volume suggests.  Client 4 experiences an even sharper correction, dropping from a FedAvg share of 18.4\% to just 8.9\%.  Conversely, Client 7 rises from 15.0\% to 23.0\%, and Client 5 from 12.9\% to 16.8\%, indicating that validators consistently rate their submissions above what dataset size alone would predict.  These shifts demonstrate that the consensus mechanism assesses submission quality rather than implicitly recovering dataset size.
The consensus weights are non-stationary: they shift as the relative quality of client submissions evolves across rounds.  Client 7, for instance, begins near its FedAvg baseline of 15.0\% but steadily climbs to 23.0\% by round 60, reflecting a sustained improvement in submission quality as judged by the validators.  Client 4 follows the opposite trajectory, declining from its initial weight toward 8.9\%.  This continuous adaptation requires no explicit scheduling or hyperparameter tuning; it emerges naturally from the evaluation scores that the consensus mechanism already produces.

These results confirm that the consensus-based weighting serves a dual purpose beyond mask selection: it provides a privacy-preserving alternative to dataset-size reporting and implicitly down-weights low-quality or adversarial contributions without requiring dedicated Byzantine-robust aggregation rules.

\subsection{Ablation Study}
\label{sec:ablation}

We ablate two key design axes of our framework: (i)~the severity of data heterogeneity, controlled by the Dirichlet concentration parameter~$\alpha$, and (ii)~the number of validators~$V$ participating in the consensus mechanism.  All experiments use CIFAR-10 with $N\!=\!20$ total nodes and are run for up to 250 communication rounds.

\begin{figure*}[t]
    \centering
    \includegraphics[width=1\linewidth]{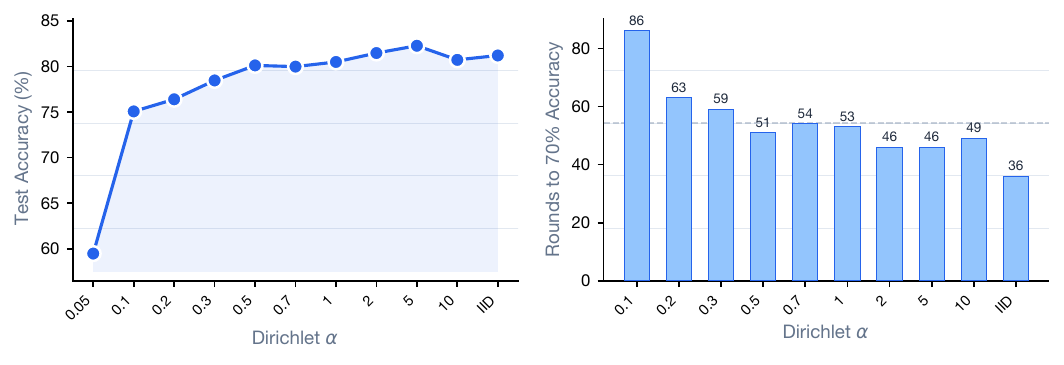}
        \caption{Effect of Dirichlet~$\alpha$ on our framework (CIFAR-10, $N\!=\!20$, $V\!=\!5$).  \emph{Left:} best test accuracy across 250 rounds.  \emph{Right:} communication rounds needed to first reach 70\% accuracy; $\alpha\!=\!0.05$ never reaches this threshold and is omitted.}
    \label{fig:ablation_alpha}
\end{figure*}

\paragraph{Effect of data heterogeneity ($\alpha$)}
Figure~\ref{fig:ablation_alpha} reports the best test accuracy achieved by our framework across eleven non-IID levels spanning $\alpha\!=\!0.05$ (extreme heterogeneity) to IID ($\alpha\!=\!10^5$), with $V\!=\!5$ validators held fixed.  Two regimes are evident.
From IID down to $\alpha\!=\!0.5$, accuracy remains within a narrow 2.2\,pp band (80.1\% to 82.3\%), indicating that the consensus mechanism effectively compensates for moderate distribution skew.  Performance begins to decline more noticeably only below $\alpha\!=\!0.3$: the $\alpha\!=\!0.1$ setting reaches 75.1\% and the most extreme case ($\alpha\!=\!0.05$) settles at 59.5\%.  Even at $\alpha\!=\!0.1$, the framework retains over 91\% of the IID-level accuracy, evidence of robustness to non-IID partitions.
The right panel of Figure~\ref{fig:ablation_alpha} shows the number of rounds required to first reach 70\% test accuracy.  Under the IID setting, the model crosses this threshold within 36 rounds; at $\alpha\!=\!0.1$ the same milestone requires 86 rounds, a 2.4$\times$ slowdown.  The $\alpha\!=\!0.05$ setting never reaches 70\% at all, confirming that it pushes the framework beyond its effective operating range.  For the moderate non-IID settings ($\alpha \geq 0.5$), convergence speed varies by fewer than ten rounds relative to IID, reinforcing the finding that the consensus mechanism absorbs much of the heterogeneity cost.

\begin{figure*}[t]
    \centering
    \includegraphics[width=1\linewidth]{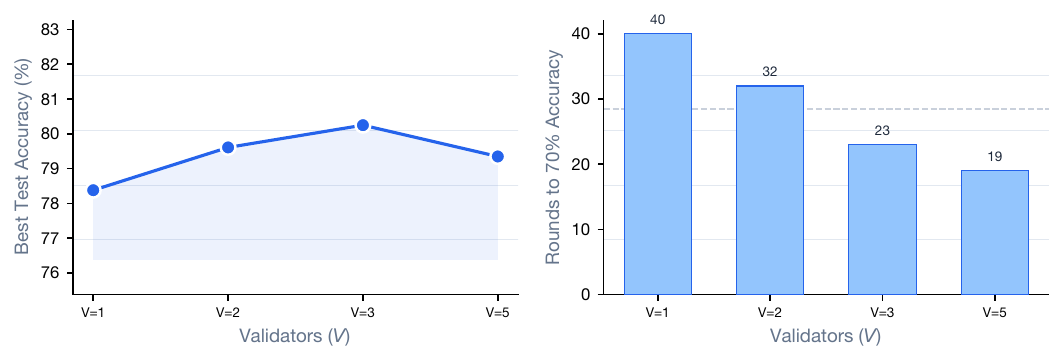}
        \caption{Effect of validator count~$V$ on our framework (CIFAR-10, $N\!=\!10$, $\alpha\!=\!0.3$).  The total training data (40\,000 samples) and validator pool (10\,000 samples) are held constant; increasing~$V$ reduces the number of training clients.  \emph{Left:}~best test accuracy.  \emph{Right:}~communication rounds to first reach 70\% accuracy.}
    \label{fig:ablation_validator}
\end{figure*}

\paragraph{Effect of validator count ($V$)}
Figure~\ref{fig:ablation_validator} isolates the effect of the number of validators on a fixed network of $N\!=\!10$ nodes with $\alpha\!=\!0.3$.  Importantly, the total data budget is held constant across all settings: 40\,000 training samples are partitioned among the $N - V$ training clients and 10\,000 samples form the validator pool, regardless of~$V$.  Increasing $V$ therefore does not change the total amount of training data; it redistributes the same data across fewer, larger client partitions while adding more evaluators to the consensus mechanism.
Best test accuracy increases from 78.4\% at $V\!=\!1$ to a peak of 80.2\% at $V\!=\!3$, then drops back to 79.3\% at $V\!=\!5$.  A single validator provides a weak consensus signal (mask selection is determined by a single evaluator's judgement, which may not generalise), while $V\!=\!5$ leaves only five training clients, reducing gradient diversity despite each client holding a larger local dataset.  The peak at $V\!=\!3$ ($V/N = 0.3$) indicates that, in this setting, a modest quorum of validators is sufficient for reliable mask selection without shrinking the training cohort too far.
The right panel of Figure~\ref{fig:ablation_validator} shows that more validators consistently accelerate early training: the 70\% accuracy milestone is reached in 40 rounds with $V\!=\!1$ but only 19 rounds with $V\!=\!5$, a 2.1$\times$ speedup.  This monotonic trend contrasts with the non-monotonic peak-accuracy curve and reflects the fact that a stronger consensus signal (more evaluators) selects better masks in early rounds when submission quality is highly variable.  In later rounds, however, the accuracy ceiling is limited by gradient diversity, which favours more training clients.

Taken together, the two ablations show that our framework is robust across a wide operating range: moderate non-IID heterogeneity ($\alpha \geq 0.5$) and modest validator quorums ($V/N \approx 0.3$) are sufficient to achieve near-peak performance, while extreme settings at either axis ($\alpha < 0.1$ or $V/N \geq 0.5$) reveal the natural limits of the consensus-based approach.

\subsection{Robustness to Gradient-Scaling Attacks}
\label{sub:scaling_robustness}

We now quantify robustness to the gradient-scaling attack defined in Section~\ref{sub:attack_models}. This attack is particularly damaging to FedAvg, where the aggregated model is a simple weighted average, so a single scaled submission can dominate the update.

We validate our solution across three axes: the four datasets, the three network topologies ($N\!=\!10,V\!=\!3$; $N\!=\!20,V\!=\!5$; $N\!=\!30,V\!=\!6$), and the three attacker fractions (10\%, 20\%, 30\%), holding $\alpha\!=\!0.3$ and the scale factor of Section~\ref{sub:attack_models} fixed. Each configuration is repeated over three seeds, for a total of 216 experiments (108 per method), and we report mean $\pm$ standard deviation.

\begin{figure*}[t]
    \centering
    \includegraphics[width=\textwidth]{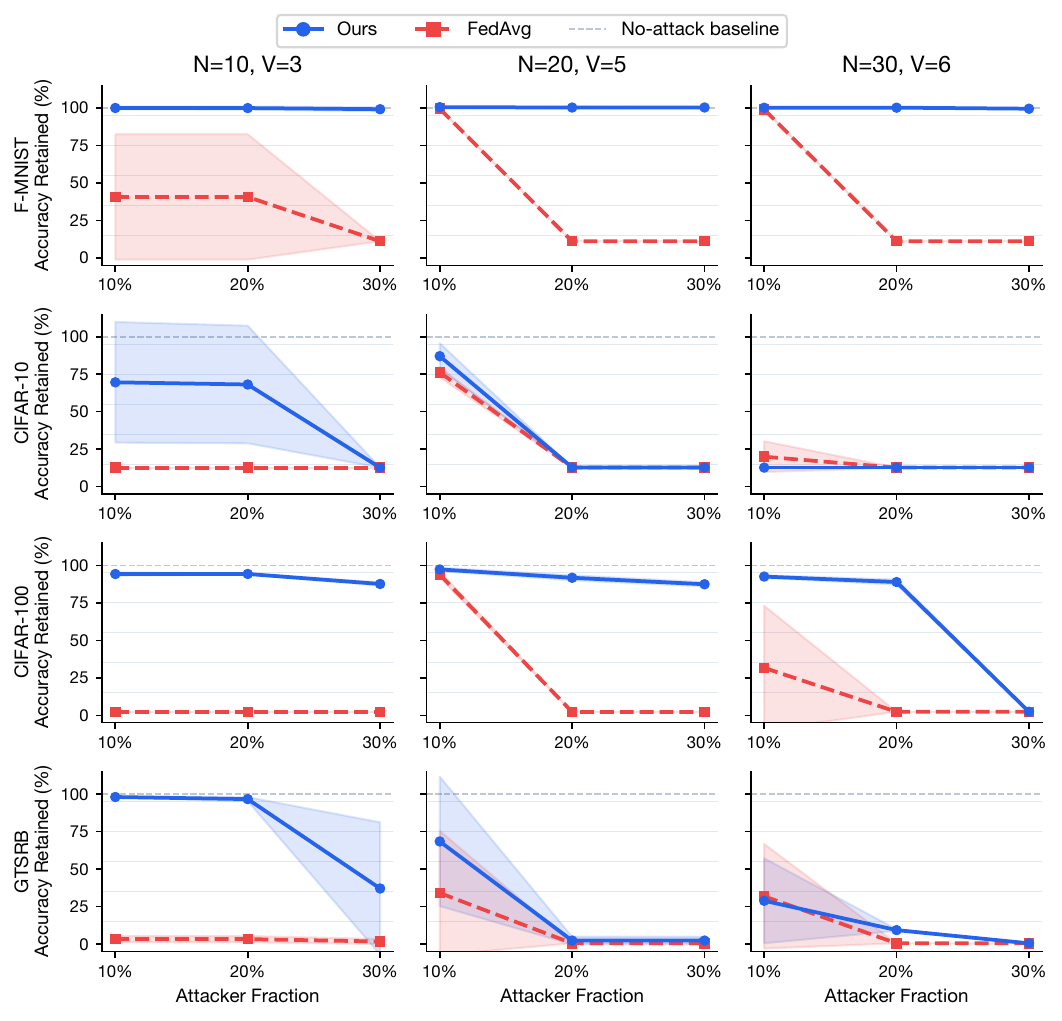}
    \caption{Accuracy retained under gradient scaling attack (scale factor 10$\times$, $\alpha\!=\!0.3$). Each subplot shows accuracy retained relative to the no-attack FedAvg baseline (dashed grey line at 100\%) as a function of attacker fraction. Rows: datasets; columns: network topologies. Shaded regions denote $\pm$1 standard deviation over three seeds. Our method (solid blue) maintains near-baseline accuracy across most configurations where FedAvg (dashed red) collapses to random chance.}
    \label{fig:scaling_retention}
\end{figure*}

Recall that accuracy retained normalizes the best accuracy under attack by the no-attack FedAvg baseline for the same topology. A value of 100\% means the attacked model matches clean FedAvg, values below 100\% quantify degradation, and values above 100\% mean the method beats the clean FedAvg baseline even under attack. Figure~\ref{fig:scaling_retention} shows these results as a $4 \times 3$ grid. A retention above 100\% therefore means that, even under attack, our method exceeds the clean FedAvg baseline, combining its clean-data edge with its resilience; the strength of the defense is seen most directly in the gap between the two curves under attack.

We highlight four findings:

\textbf{(1) Near-complete resilience on F-MNIST.}\quad  Our method retains $\geq$99.5\% of clean accuracy across all topologies and attacker fractions. At $N\!=\!20,V\!=\!5$, it achieves 100.6\% retention even at 10\% attackers, indicating that the mask-consensus mechanism fully neutralizes the scaled updates. FedAvg, by contrast, collapses to random chance (11.1\% retention) at $\geq$20\% attackers for the same topology, representing a complete loss of learned representations.

\textbf{(2) Strong protection on CIFAR-100.}\quad  At $N\!=\!10,V\!=\!3$, our method retains 94.1\%, 94.2\%, and 87.5\% of baseline accuracy at 10\%, 20\%, and 30\% attacker fractions respectively, while FedAvg collapses to 2.1\% retention (random chance for 100 classes) at every fraction tested. The pattern is consistent at $N\!=\!20,V\!=\!5$, where our method retains 96.8\% at 10\% attackers and degrades gracefully to 87.0\% at 30\%, whereas FedAvg collapses completely at $\geq$20\% attackers. This demonstrates that the saliency-based masking is particularly effective on fine-grained classification tasks where FedAvg is most vulnerable.

\textbf{(3) A collapse boundary shifted later.}\quad  Our method holds accuracy at attacker fractions where FedAvg has already collapsed to random chance, shifting the point of collapse at least one fraction step later. On CIFAR-100 at \mbox{$N\!=\!10,V\!=\!3$} the decline is gradual, retaining 94.1\%, 94.2\%, and 87.5\% from 10\% to 30\% attackers while FedAvg sits at 2.1\% throughout. On GTSRB at \mbox{$N\!=\!10,V\!=\!3$} our method holds 97.6\% at 10\% and 96.2\% at 20\% attackers, full strength at a point where FedAvg has already fallen to 3.3\%, before declining to 36.9\% at 30\%. That final drop is sharper than on CIFAR-100, but across these datasets the protected envelope still extends at least one attacker-fraction step beyond FedAvg.

\textbf{(4) Both methods share ultimate limits.}\quad  At sufficiently high attacker fractions, both methods eventually collapse. On CIFAR-10 at $N\!=\!20,V\!=\!5$, both methods collapse to random chance at $\geq$20\% attackers; on CIFAR-100 at $N\!=\!30,V\!=\!6$, both collapse at 30\%. This is expected: when the fraction of poisoned gradients exceeds the Byzantine fault tolerance threshold of the network, no aggregation-level defense can fully recover. The critical observation is that our method consistently shifts this collapse boundary by one or more fraction steps relative to FedAvg, extending the safe operating envelope of the system.

\paragraph{Mechanism.}
The mask-consensus mechanism provides two complementary layers of defense. First, the saliency-based mask selects model coordinates based on their relevance to the validation loss rather than their raw magnitude. Since gradient scaling uniformly inflates all coordinates, the mask filters out disproportionately amplified parameters that do not align with the validation signal. Second, the score-weighted aggregation assigns lower weights to submissions that degrade validator-assessed accuracy, limiting the influence of scaled updates even when the mask does not fully suppress them. Together, these mechanisms explain the observed robustness: the attack amplifies magnitude without improving validation-loss relevance, so the mask naturally attenuates the poisoned coordinates.

\subsection{Label-Free Composition of an IID Validator Pool}
\label{subsec:dqs_results}

Our framework does not require any individual validator to hold balanced data. Validators may each be arbitrarily class-skewed; what matters is that the admitted validator set is \emph{collectively} IID, the \emph{pool-iid} setting of Section~\ref{sub:experiment_performance}. The admission attestation makes this achievable without labels: from the reference model's public output distributions it derives, for each validator, a class profile (the mean softmax it produces over its private samples), and the network uses these profiles to select validators whose combined test set is balanced.

On the FashionMNIST reference model, we draw a population of 60 validators whose dominant classes follow a skewed prior, so the population itself is imbalanced (some classes common, some rare) and a validator set chosen at random is not IID. Each validator holds 120 samples. We then select validators to make their combined test set uniform over the ten classes, comparing three strategies: \emph{random} selection; \emph{score}, a greedy selection driven by the label-free class profile; and \emph{oracle}, the same greedy selection driven by the true (hidden) labels. For every strategy we measure how IID the resulting test set turns out to be, evaluated on the true labels, as the total-variation distance of its class distribution from uniform; lower is more IID.

\begin{figure*}[t]
  \centering
  \includegraphics[width=0.92\linewidth]{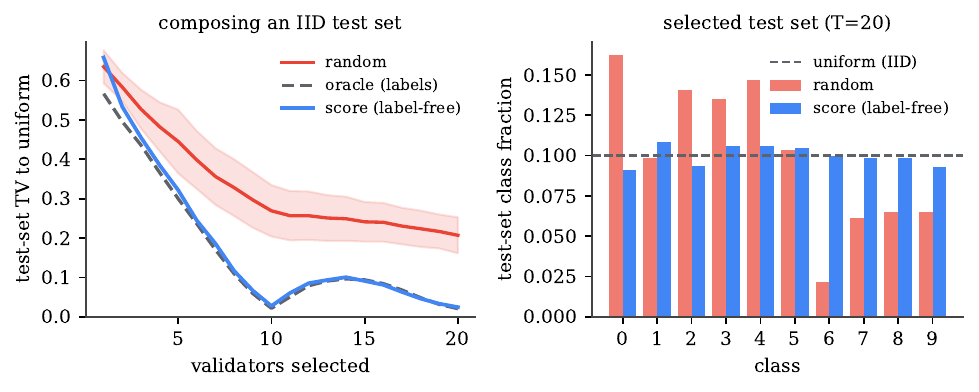}
  \caption{Label-free composition of an IID validator pool (FashionMNIST, population of 60 skewed validators). \emph{Left:} as validators are selected, the combined test set's distance from a uniform (IID) class distribution; the label-free score (blue) tracks the label-based oracle (dashed) almost exactly and far outperforms random selection (red). \emph{Right:} the resulting test-set class distribution at twenty validators is flat under the score but lopsided under random, which over-represents common classes and starves rare ones.}
  \label{fig:dqs}
\end{figure*}

The label-free profile is accurate (mean total-variation distance $0.08$ from the true class distribution), and it composes an IID test set as well as the oracle (Table~\ref{tab:dqs}, Figure~\ref{fig:dqs}). The population aggregate sits at total variation $0.16$ from uniform, and random selection of twenty validators stays at $0.21$: picking validators blindly does not yield a balanced set. Selecting with the label-free score instead drives the combined test set to total variation $0.025$ from uniform, indistinguishable from the oracle's $0.021$ and almost an order of magnitude closer to IID than random. The two greedy curves in Figure~\ref{fig:dqs} (left) track each other throughout, and the resulting class histogram (right) is flat under the score where random over-represents the common classes and starves the rare ones.

In short, a privacy-preserving score over public output distributions is enough to read each validator's class makeup and assemble a validator pool that is IID in aggregate, without inspecting raw data or labels and without requiring any validator to be balanced on its own.

\begin{table}[t]
\centering
\caption{Composing an IID validator test set on FashionMNIST (population of 60 individually-skewed validators, 20 selected). TV is the total-variation distance of the selected test set's class distribution from uniform; lower is more IID. The label-free score matches the label-based oracle and far outperforms random selection.}
\label{tab:dqs}
\small
\resizebox{\columnwidth}{!}{\begin{tabular}{lc}
\toprule
Selection strategy & Test-set TV to uniform \\
\midrule
Imbalanced population (all 60)   & $0.160$ \\
Random (20 validators)           & $0.207$ \\
Oracle, true labels (20)         & $0.021$ \\
Score, label-free (20)           & $\mathbf{0.025}$ \\
\bottomrule
\end{tabular}}
\end{table}

\subsection{Cost of Zero-Knowledge Admission and Mask Auditing}
\label{subsec:zk_results}
We now quantify the overhead of the cryptographic machinery that admits validators and audits their masks. Because the attestation runs once per validator at admission, its cost is amortized over every subsequent aggregation round. We benchmark each reference model on a single GPU server and report four quantities: the cost of proving and verifying admission, the accuracy preserved under circuit quantization, the cost and privacy of the challenge audit, and the cost of one complete admission round. Unless noted, circuits use a fixed-point scale of 13 and a circuit-size cap of $k=16$ logrows.

\begin{table}[t]
\centering
\caption{Per-phase cost of the framework, asymptotic and measured in one place. The time column is recalled from Table~\ref{tab:complexity}, where the full space and communication bounds are derived, and is paired here with the measured cost on the FashionMNIST reference model. Admission is paid once per validator and amortized over all later rounds; evaluation, masking, and consensus recur each round; the challenge runs per audit. An entry of \emph{none} marks a phase with no cryptographic overhead, the standard forward and backward passes or Byzantine Fault Tolerant messaging, which we do not benchmark separately. One-time admission artifacts are 97~s setup, an 18.4~GB proving key released after admission, and an 8.6~MB verifying key. Symbols follow Table~\ref{tab:complexity}: $b$ batch size, $F$ a reference-model forward pass, $M$ pool size, $n$ test-set size, $d$ parameters, $V$ validators, $P$ challenge peers.}
\label{tab:cost}
\resizebox{\columnwidth}{!}{\begin{tabular}{llll}
\toprule
Phase (cadence) & Time & Measured & Footprint \\
\midrule
Admission (once / validator)     & $\tilde{O}(bF)$       & 84~s prove, 3.0~s verify & 60.5~GB host, 879~KB proof \\
Eval + masking (per round)       & $O(M(nF{+}d))$        & none                     & $O(d)$ state, $O(Md)$ comm. \\
Challenge audit (per challenge)  & $O(P(bF{+}d))$        & 69~ms                    & 12.5~KB exposed \\
Consensus (per round)            & $O(Vd)$               & none                     & $O(V^2)$ messages \\
\bottomrule
\end{tabular}}
\end{table}

\subsubsection{Proof Generation and Verification Cost}
Table~\ref{tab:cost} pairs the asymptotic cost of each protocol phase with the measured wall-clock for the FashionMNIST reference model, and the one-time admission proof dominates, exactly as the analysis of Section~\ref{subsec:complexity} predicts. Admission is the only phase whose cost is superlinear in the reference model, at \mbox{$\tilde{O}(bF)$} proving time and \mbox{$O(bF)$} space, whereas verification is succinct, constant in the circuit size, and the recurring per-round phases of evaluation, masking, and consensus carry no cryptographic proving cost. The measured times follow this ordering directly. Proving takes 84~seconds and verification 3.0~seconds, producing an 879~KB proof, while the one-time setup takes 97~seconds and yields an 18.4~GB proving key that the validator discards once admitted. The peak host memory is 60.5~GB. Constant-size verification keeps the 3.0-second check cheap enough to run on-chain in the \texttt{Verifier} contract, while proving is borne once by the candidate at admission and amortized over every later round. Because both the proving time and its \mbox{$O(bF)$} space grow with the cost of the reference-model forward pass, scaling this attestation to larger reference models is the main practical constraint of the framework, which we examine in Section~\ref{sec:limitations}.

\subsubsection{Quantization and Attestation Accuracy}
A zero-knowledge circuit computes over a fixed-point field, so the attested computation could in principle drift from the floating-point reference. It does not. At a scale of 13, the circuit reproduces the reference model's output logits with a mean absolute error of $3.1\times10^{-3}$ and a mean absolute percentage error of 0.066\%, the largest single-logit error being 0.014. These deviations are far smaller than the gap between competing class logits, so the quantized model's top-1 predictions match the floating-point model and the 92.7\% FashionMNIST test accuracy is preserved. The attestation therefore certifies the same decisions the validator would make in floating point, which is what the data-quality check depends on.

\subsubsection{Challenge Protocol Cost and Privacy}
The challenge audit must do two things at once: catch a validator whose mask was not computed on its attested data, and avoid leaking that data to the peers who check it. We evaluate both.

Detection is exact and cheap. Over 36 challenges per validator type, an honest validator's mask matches the peer-reconstructed gradient at an agreement of 1.000 and is always accepted, while a fabricated mask agrees at 0.08 and is always rejected, against an acceptance threshold of 0.95. The exposed activations range from 12~KB to 2~MB depending on batch size and cut layer, and a peer completes its check in a few milliseconds, so the audit adds negligible cost to a round.

Privacy depends on where the model is cut. Following the model-inversion attack on split inference of He et al.~\cite{he2019model}, we train a decoder network of transposed convolutions to reconstruct a validator's inputs from the exposed activations, and score the reconstruction with SSIM, for which lower is more private. Table~\ref{tab:cut_privacy} sweeps the cut layer for each model. A deeper, more downsampled cut produces a more abstract activation that is harder to invert: on CIFAR-10 the reconstruction SSIM falls from 0.88 at \texttt{layer2} to 0.49 at \texttt{layer3\_pool}, and on CIFAR-100 from 0.96 to 0.53 at \texttt{layer4}. The audit bounds how deep the cut can go, because it reconstructs gradients only for parameters after the cut: once too few remain, honest agreement drops below the acceptance threshold, the rows marked \emph{weak} in Table~\ref{tab:cut_privacy}. The rule we follow is to take the deepest cut that still keeps a sound audit, which gives an SSIM of 0.49 on CIFAR-10, 0.70 on GTSRB, and 0.53 on CIFAR-100. FashionMNIST is the exception, with an SSIM of 0.88 at its only available cut, a shallow-model limitation we return to in Section~\ref{sec:limitations}.

We also tested adding Gaussian noise to the exposed activations and found it ineffective here. An inverter trained on noisy activations learns to average the noise away, so the SSIM barely moves, while the same noise corrupts the reconstructed gradient and collapses the audit. Cut depth, not noise, is the control that matters. A validator that needs a stronger guarantee can additionally apply a dedicated activation-leakage-reduction method~\cite{cihangiroglu2026lightsplit} to the activations it exposes, lowering reconstructability without changing the cut. In return, the validator must apply the same method at test time, so that its exposed activations stay consistent with the outputs the peers recompute during the audit.

\begin{table}[htbp]
\centering
\caption{Cut-layer selection trades validator data privacy against audit strength. SSIM scores how well an inverter reconstructs the input from the exposed activations (lower is more private). \emph{Honest} and \emph{Adv.} are the mask-agreement rates of an honest and a fabricated validator; the audit is sound when honest agreement stays above the 0.95 acceptance threshold and adversarial agreement stays low. \emph{weak} marks a cut too deep to audit reliably. Bold marks the cut used in our framework.}
\label{tab:cut_privacy}
\resizebox{\columnwidth}{!}{\begin{tabular}{llrrrl}
\toprule
Dataset & Cut layer & SSIM & Honest & Adv. & Audit \\
\midrule
FashionMNIST & \textbf{features} & 0.876 & 1.000 & 0.080 & ok \\
\midrule
CIFAR-10 & layer2 & 0.876 & 1.000 & 0.076 & ok \\
 & \textbf{layer3\_pool} & 0.494 & 1.000 & 0.080 & ok \\
 & layer3\_res & 0.288 & 0.762 & 0.324 & weak \\
\midrule
GTSRB & layer2 & 0.914 & 1.000 & 0.065 & ok \\
 & \textbf{layer3\_pool} & 0.700 & 1.000 & 0.082 & ok \\
 & layer3\_res & 0.588 & 0.868 & 0.189 & weak \\
\midrule
CIFAR-100 & layer2 & 0.963 & 1.000 & 0.092 & ok \\
 & layer3 & 0.873 & 1.000 & 0.074 & ok \\
 & \textbf{layer4} & 0.532 & 1.000 & 0.174 & ok \\
\bottomrule
\end{tabular}}
\end{table}

\subsubsection{Overhead Analysis}
Finally, we run one complete admission round on FashionMNIST with three validators. Each candidate proves correct execution of the reference model on its private data and is admitted after on-chain proof verification and an off-chain data-quality score of $Q\approx1.0$, and all three are admitted. Admission is dominated by proving, at 84~seconds per validator on average, with 3~second verification and an 879~KB proof, while the circuit setup is a one-time cost shared across validators. A subsequent challenge of an admitted validator completes in 69~milliseconds at an agreement of 1.000, exposing 12.5~KB of activations. The round confirms that the mechanisms compose as intended: the dominant cost is the one-time zero-knowledge admission proof, and the recurring challenge audit is effectively free.

\section{Scalability and Limitations}
\label{sec:limitations}

The attestation cost in Section~\ref{subsec:zk_results} is reported for the compact FashionMNIST reference model, the setting in which the full pipeline runs on a single machine. The memory needed to generate a HALO2 proof grows with the size of the reference model, and on our hardware the larger reference models, the 6.6M-parameter ResNet-9 and the 11.2M-parameter ResNet-18, exceed the memory available for proving. This bounds the size of the reference model that can be attested on a single commodity server today, a property of current proving infrastructure rather than of the protocol, since the attestation is a one-time, per-validator, and embarrassingly parallel cost (Section~\ref{subsec:complexity}). Three directions reduce the requirement. First, the model used for attestation need not be the model being trained: a small fixed probe network, large enough to expose data-quality statistics but no larger, bounds the circuit independently of task complexity. Second, circuit decomposition with recursive proof aggregation can attest a large model in memory-bounded pieces. Third, the proof can be offloaded to higher-memory or GPU-backed provers. We take a smaller fixed reference probe to be the most direct lever and leave it to future work.

The privacy of the challenge audit depends on the depth of the cut layer, which in turn depends on the reference model. A deep model admits a deep, downsampled cut whose activations resist reconstruction, reaching an SSIM near 0.5 on the ResNet datasets, whereas a shallow model admits only a shallow cut that leaks more: the single-block SimpleCNN on FashionMNIST reaches an SSIM of 0.88. This is a property of small reference models rather than of the audit, and a deeper reference model improves both privacy and the scaling discussed above.

A validator scores each submission on its private test set, so a dishonest validator could draw that set from data it had already trained on, judging models on samples it has already seen and biasing the resulting data-quality scores and consensus weights in its favor. The admission attestation does not preclude this: it certifies that the validator executed the reference model correctly on its committed data, not that the data was withheld from training. We acknowledge this as a residual threat. Closing it requires an additional zero-knowledge proof at admission, in which the validator attests that its evaluation set is disjoint from its training set. We treat this proof as a deployment-time safeguard rather than part of our evaluation, since it is orthogonal to our admission attestation and adds a further circuit to the admission cost.

\section{Related Work}
\label{sec:related}
Our framework sits where three research lines meet: aggregation methods that cope with heterogeneous or adversarial clients, decentralized learning that removes the central server, and verifiable computation that lets a party prove what it did without exposing its data. We review each line below and mark the same gap throughout. None of them treats data quality as a property a validator must \emph{prove}, in zero knowledge and in private, before it is allowed to shape aggregation. Making that proof the entry condition is what sets our robust and privacy-preserving Swarm Learning framework apart.

\subsection{Robust Aggregation under Heterogeneity and Adversaries}

One line adapts the aggregation rule to non-IID data. FedProx~\cite{li2020federated} adds a proximal term that keeps each local update near the global model, FedNova~\cite{wang2020tackling} normalizes updates by the number of local steps so that clients training longer do not dominate, and SCAFFOLD~\cite{karimireddy2020scaffold} carries control variates that correct client drift. These methods steady convergence under skew, yet they retain the dataset-size weighting of FedAvg and, for FedNova and SCAFFOLD, exchange extra statistics about each client's local training. That metadata raises communication cost and widens the privacy surface, and all three assume the clients reporting it behave honestly.

A second line hardens aggregation against malicious updates. Trimmed Mean~\cite{yin2018byzantine} discards the most extreme value of each coordinate before averaging, Krum~\cite{blanchard2017machine} selects the update closest to its neighbors in $L_2$ distance, Bulyan~\cite{guerraoui2018hidden} applies a coordinate-wise trim on top of repeated Krum selection, and Flame~\cite{nguyen2022flame} clusters updates by cosine similarity, then clips and adds noise to suppress stealthy poisoning. Each rule treats an adversary as a statistical outlier, an assumption that weakens under non-IID data, where honest updates are themselves diverse, and against poisoning crafted to stay inside the benign range~\cite{fang2020local}. More fundamentally, every method in both lines runs at a trusted aggregator and inspects only the submitted weights. None verifies the data on which a participant trained or evaluated, which is exactly the surface a decentralized protocol exposes.

\subsection{Decentralized and Swarm Learning}

Swarm Learning~\cite{warnat2021swarm} removes the central server and coordinates training across a peer network through blockchain smart contracts, training clinical models across institutions without sharing patient records. Other decentralized schemes reach agreement through peer exchange rather than a server, for instance by mutual knowledge transfer between neighbors~\cite{li2021decentralized}. Removing the server removes the single point of failure, but it also removes the one party trusted to judge an update and hands that judgment to peers whose own data is never checked. These protocols coordinate \emph{who} aggregates and \emph{when}, and leave \emph{whether a peer's data is sound} outside the protocol. Our framework brings that question inside it.

\subsection{Verifiable Computation and Privacy-Preserving Validation}

Cryptographic tools let a party prove that a computation was carried out correctly, or keep its inputs private, with no trusted observer. Zero-knowledge proofs~\cite{goldwasser2019knowledge}, realized in systems such as HALO2~\cite{halo2} with tooling like EZKL~\cite{ezkl}, can attest that a model ran on committed inputs while revealing nothing about them. A growing body of work applies proof systems to machine learning, both to certify that training or inference was performed as claimed and to bring verifiable computation and trusted sharing into federated and swarm pipelines~\cite{arikkat2025sectis,arazzi2025privacy}. On the privacy side, secure aggregation~\cite{bonawitz2017practical} hides individual updates from the aggregator, and Split Learning~\cite{vepakomma2018split} exchanges intermediate activations at a cut layer in place of raw data. These mechanisms verify \emph{how} a computation ran, or hide \emph{what} entered it, yet none turns the result into an admission test on a participant's data. Our framework makes that test the entry point: a candidate proves in zero knowledge that the reference model ran correctly on its private set, and the attested outputs must clear a data-quality threshold before the validator can shape aggregation. The same cut-layer principle then supports a low-cost re-audit of a validator's mask, avoiding a proof over the full training pass.

\subsection{Data Quality and Contribution Valuation}

A final line weights or selects clients by an estimate of their usefulness rather than their dataset size. Precision-weighted aggregation~\cite{reyes2025precision} scales each contribution by a confidence estimate, and contribution-valuation methods score clients by their marginal effect on model quality. These signals are computed by a trusted coordinator from disclosed updates or reported metadata, and they rank participants already in the protocol rather than gate entry to it. We instead derive each aggregation weight from validator consensus over private evaluations, and we make a proven data-quality score the condition for becoming a validator at all.

Across these lines, robustness is pursued at the aggregation rule, decentralization at the coordination layer, and privacy at the message level, while data quality is assumed rather than verified. By making a verifiable, privacy-preserving data-quality test a first-class admission criterion under BFT consensus, our framework turns that assumption into a checked property, which is what lets it hold accuracy under the gradient-scaling attack that drives FedAvg to random chance (Section~\ref{sub:scaling_robustness}).

\section{Conclusion}
\label{sec:conclusion}

We presented a robust and privacy-preserving Swarm Learning framework that makes zero-knowledge validator admission the condition for shaping the global model. Three mechanisms operate under one Byzantine Fault Tolerant validator workflow. A zero-knowledge attestation admits a validator only after it proves, without revealing a single sample, that a reference model ran correctly on its private data and that the result clears a quality threshold. A quality-aware mask consensus then sets each aggregation weight from validator scores rather than dataset size. A challenge-based audit, adapted from Split Learning, re-checks a validator's mask without a proof over the full training pass. Model proposals, admission proofs, and consensus decisions become first-class, auditable objects on chain, which allows the protocol to avoid the cost of end-to-end zero-knowledge training.

Empirically, the framework matches FedAvg on clean data across four benchmarks, with its clearest gains on the most heterogeneous partitions. Under a gradient-scaling attack that collapses FedAvg to random chance, the framework retains 94\% of its clean accuracy on CIFAR-100 and more than 99\% on FashionMNIST, where FedAvg holds only 2.1\% and 11\%, and it gradually degrades rather than collapsing. This gradient-scaling result is the resilience that we demonstrate. Resistance to broader Byzantine behavior is a goal of the attestation, challenge, and BFT layers rather than a demonstrated result, and we leave that evaluation to future work. Ablations mark the operating envelope: the consensus absorbs moderate heterogeneity ($\alpha \geq 0.5$) and modest validator quorums ($V/N \approx 0.3$) with little loss, while extreme heterogeneity ($\alpha < 0.1$) or large quorums ($V/N \geq 0.5$) erode accuracy or gradient diversity, and, like any aggregation-level defense, the protocol cannot recover once the poisoned fraction exceeds the network's Byzantine tolerance.

By linking privacy, adversarial resilience, and on-chain accountability to a verifiable admission check rather than to client-reported metadata or a trusted coordinator, the framework closes the validator-side attack surface that decentralization would otherwise leave open. Future work will carry out the multi-attack evaluation and calibrate the data-quality threshold against downstream accuracy. We further plan to refine the challenge protocol for gradient masking, explore additional data-availability back ends, and extend the validator logic to multi-modal and streaming data.

\section*{CRediT Authorship Contribution Statement}
\begingroup
\raggedright
\begin{itemize}
    \item \textbf{Marco Arazzi:} Conceptualization, Methodology, Writing~--~review \& editing.
    \item \textbf{Mert Cihangiroglu:} Conceptualization, Data curation, Formal analysis, Investigation, Methodology, Software, Validation, Visualization, Writing~--~original draft, Writing~--~review \& editing.
    \item \textbf{Antonino Nocera:} Conceptualization, Methodology, Supervision, Writing~--~review \& editing.
\end{itemize}
\endgroup

\section*{Use of AI Assistants}
We used generative AI tools to polish author-written text, assist with parts of the experimental and analysis code, and help produce figures from our results. All AI-assisted work was reviewed by the authors, who take full responsibility for the final text, claims, code, and figures.

{\footnotesize
\bibliography{biblio}

@inproceedings{mcmahan2017communication,
  title={Communication-efficient learning of deep networks from decentralized data},
  author={McMahan, Brendan and Moore, Eider and Ramage, Daniel and Hampson, Seth and y Arcas, Blaise Aguera},
  booktitle={Artificial intelligence and statistics},
  pages={1273--1282},
  year={2017},
  organization={Pmlr}
}

@article{konevcny2016federated,
  title={Federated optimization: Distributed machine learning for on-device intelligence},
  author={Kone{\v{c}}n{\`y}, Jakub and McMahan, H Brendan and Ramage, Daniel and Richt{\'a}rik, Peter},
  journal={arXiv preprint arXiv:1610.02527},
  year={2016}
}

@inproceedings{bonawitz2017practical,
  title={Practical secure aggregation for privacy-preserving machine learning},
  author={Bonawitz, Keith and Ivanov, Vladimir and Kreuter, Ben and Marcedone, Antonio and McMahan, H Brendan and Patel, Sarvar and Ramage, Daniel and Segal, Aaron and Seth, Karn},
  booktitle={proceedings of the 2017 ACM SIGSAC Conference on Computer and Communications Security},
  pages={1175--1191},
  year={2017}
}

@article{bonawitz2019towards,
  title={Towards federated learning at scale: System design},
  author={Bonawitz, Keith and Eichner, Hubert and Grieskamp, Wolfgang and Huba, Dzmitry and Ingerman, Alex and Ivanov, Vladimir and Kiddon, Chloe and Kone{\v{c}}n{\`y}, Jakub and Mazzocchi, Stefano and McMahan, Brendan and others},
  journal={Proceedings of machine learning and systems},
  volume={1},
  pages={374--388},
  year={2019}
}

@article{li2020challenges,
  title={Federated learning: Challenges, methods, and future directions},
  author={Li, Tian and Sahu, Anit Kumar and Talwalkar, Ameet and Smith, Virginia},
  journal={IEEE signal processing magazine},
  volume={37},
  number={3},
  pages={50--60},
  year={2020},
  publisher={IEEE}
}

@article{arikkat2025sectis,
  title={Sectis: A framework to secure cti sharing},
  author={Arikkat, Dincy R and Cihangiroglu, Mert and Conti, Mauro and KA, Rafidha Rehiman and Nicolazzo, Serena and Nocera, Antonino and others},
  journal={Future Generation Computer Systems},
  volume={164},
  pages={107562},
  year={2025},
  publisher={Elsevier}
}

@article{arazzi2025privacy,
  title={Privacy Preserving and Robust Aggregation for Cross-Silo Federated Learning in Non-IID Settings},
  author={Arazzi, Marco and Cihangiroglu, Mert and Nocera, Antonino},
  journal={arXiv preprint arXiv:2503.04451},
  year={2025}
}

@misc{ezkl,
  author       = {ZKonduit Inc.},
  title        = {EZKL: Easy Zero-Knowledge Inference Library},
  howpublished = {\url{https://ezkl.xyz/}},
  year         = {2024}
}

@article{warnat2021swarm,
  title={Swarm learning for decentralized and confidential clinical machine learning},
  author={Warnat-Herresthal, Stefanie and Schultze, Hartmut and Shastry, Krishnaprasad Lingadahalli and Manamohan, Sathyanarayanan and Mukherjee, Saikat and Garg, Vishesh and Sarveswara, Ravi and H{\"a}ndler, Kristian and Pickkers, Peter and Aziz, N Ahmad and others},
  journal={Nature},
  volume={594},
  number={7862},
  pages={265--270},
  year={2021},
  publisher={Nature Publishing Group UK London}
}

@incollection{goldwasser2019knowledge,
  title={The knowledge complexity of interactive proof-systems},
  author={Goldwasser, Shafi and Micali, Silvio and Rackoff, Chales},
  booktitle={Providing sound foundations for cryptography: On the work of shafi goldwasser and silvio micali},
  pages={203--225},
  year={2019}
}

@inproceedings{castro1999practical,
  title={Practical byzantine fault tolerance},
  author={Castro, Miguel and Liskov, Barbara and others},
  booktitle={OsDI},
  volume={99},
  number={1999},
  pages={173--186},
  year={1999}
}

@phdthesis{buchman2016tendermint,
  title={Tendermint: Byzantine fault tolerance in the age of blockchains},
  author={Buchman, Ethan},
  year={2016},
  school={University of Guelph}
}

@inproceedings{yao1982protocols,
  title={Protocols for secure computations},
  author={Yao, Andrew C},
  booktitle={23rd annual symposium on foundations of computer science (sfcs 1982)},
  pages={160--164},
  year={1982},
  organization={IEEE}
}

@inproceedings{groth2016size,
  title={On the size of pairing-based non-interactive arguments},
  author={Groth, Jens},
  booktitle={Annual international conference on the theory and applications of cryptographic techniques},
  pages={305--326},
  year={2016},
  organization={Springer}
}

@article{wood2014ethereum,
  title={Ethereum: A secure decentralised generalised transaction ledger},
  author={Wood, Gavin and others},
  journal={Ethereum project yellow paper},
  volume={151},
  number={2014},
  pages={1--32},
  year={2014}
}

@inproceedings{kate2010constant,
  title={Constant-size commitments to polynomials and their applications},
  author={Kate, Aniket and Zaverucha, Gregory M and Goldberg, Ian},
  booktitle={International conference on the theory and application of cryptology and information security},
  pages={177--194},
  year={2010},
  organization={Springer}
}

@article{li2020federated,
  title={Federated optimization in heterogeneous networks},
  author={Li, Tian and Sahu, Anit Kumar and Zaheer, Manzil and Sanjabi, Maziar and Talwalkar, Ameet and Smith, Virginia},
  journal={Proceedings of Machine learning and systems},
  volume={2},
  pages={429--450},
  year={2020}
}

@article{wang2020tackling,
  title={Tackling the objective inconsistency problem in heterogeneous federated optimization},
  author={Wang, Jianyu and Liu, Qinghua and Liang, Hao and Joshi, Gauri and Poor, H Vincent},
  journal={Advances in neural information processing systems},
  volume={33},
  pages={7611--7623},
  year={2020}
}

@inproceedings{karimireddy2020scaffold,
  title={Scaffold: Stochastic controlled averaging for federated learning},
  author={Karimireddy, Sai Praneeth and Kale, Satyen and Mohri, Mehryar and Reddi, Sashank and Stich, Sebastian and Suresh, Ananda Theertha},
  booktitle={International conference on machine learning},
  pages={5132--5143},
  year={2020},
  organization={PMLR}
}

@article{blanchard2017machine,
  title={Machine learning with adversaries: Byzantine tolerant gradient descent},
  author={Blanchard, Peva and El Mhamdi, El Mahdi and Guerraoui, Rachid and Stainer, Julien},
  journal={Advances in neural information processing systems},
  volume={30},
  year={2017}
}

@inproceedings{yin2018byzantine,
  title={Byzantine-robust distributed learning: Towards optimal statistical rates},
  author={Yin, Dong and Chen, Yudong and Kannan, Ramchandran and Bartlett, Peter},
  booktitle={International conference on machine learning},
  pages={5650--5659},
  year={2018},
  organization={Pmlr}
}

@inproceedings{guerraoui2018hidden,
  title={The hidden vulnerability of distributed learning in byzantium},
  author={Guerraoui, Rachid and Rouault, S{\'e}bastien and others},
  booktitle={International conference on machine learning},
  pages={3521--3530},
  year={2018},
  organization={PMLR}
}

@inproceedings{nguyen2022flame,
  title={$\{$FLAME$\}$: Taming backdoors in federated learning},
  author={Nguyen, Thien Duc and Rieger, Phillip and Chen, Huili and Yalame, Hossein and M{\"o}llering, Helen and Fereidooni, Hossein and Marchal, Samuel and Miettinen, Markus and Mirhoseini, Azalia and Zeitouni, Shaza and others},
  booktitle={31st USENIX security symposium (USENIX Security 22)},
  pages={1415--1432},
  year={2022}
}

@inproceedings{fang2020local,
  title={Local model poisoning attacks to $\{$Byzantine-Robust$\}$ federated learning},
  author={Fang, Minghong and Cao, Xiaoyu and Jia, Jinyuan and Gong, Neil},
  booktitle={29th USENIX security symposium (USENIX Security 20)},
  pages={1605--1622},
  year={2020}
}

@misc{halo2,
  author       = {Bowe, Sean and Grigg, Jack and Hopwood, Daira},
  title        = {The {halo2} Zero-Knowledge Proving System},
  howpublished = {\url{https://zcash.github.io/halo2/}},
  year         = {2022}
}

@inproceedings{grassi2021poseidon,
  title={Poseidon: A new hash function for $\{$Zero-Knowledge$\}$ proof systems},
  author={Grassi, Lorenzo and Khovratovich, Dmitry and Rechberger, Christian and Roy, Arnab and Schofnegger, Markus},
  booktitle={30th USENIX Security Symposium (USENIX Security 21)},
  pages={519--535},
  year={2021}
}

@article{krizhevsky2009learning,
  title={Learning multiple layers of features from tiny images},
  author={Krizhevsky, Alex and Hinton, Geoffrey and others},
  year={2009},
  publisher={Toronto, ON, Canada}
}

@article{xiao2017fashion,
  title={Fashion-mnist: a novel image dataset for benchmarking machine learning algorithms},
  author={Xiao, Han and Rasul, Kashif and Vollgraf, Roland},
  journal={arXiv preprint arXiv:1708.07747},
  year={2017}
}

@article{stallkamp2012man,
  title={Man vs. computer: Benchmarking machine learning algorithms for traffic sign recognition},
  author={Stallkamp, Johannes and Schlipsing, Marc and Salmen, Jan and Igel, Christian},
  journal={Neural networks},
  volume={32},
  pages={323--332},
  year={2012},
  publisher={Elsevier}
}

@inproceedings{xie2019dba,
  title={Dba: Distributed backdoor attacks against federated learning},
  author={Xie, Chulin and Huang, Keli and Chen, Pin-Yu and Li, Bo},
  booktitle={International conference on learning representations},
  year={2019}
}

@inproceedings{walter2024mitigating,
  title={Mitigating distributed backdoor attack in federated learning through mode connectivity},
  author={Walter, Kane and Mohammady, Meisam and Nepal, Surya and Kanhere, Salil S},
  booktitle={Proceedings of the 19th ACM Asia Conference on Computer and Communications Security},
  pages={1287--1298},
  year={2024}
}

@article{li2021decentralized,
  title={Decentralized federated learning via mutual knowledge transfer},
  author={Li, Chengxi and Li, Gang and Varshney, Pramod K},
  journal={IEEE Internet of Things Journal},
  volume={9},
  number={2},
  pages={1136--1147},
  year={2021},
  publisher={IEEE}
}

@article{reyes2025precision,
  title={Precision-Weighted Federated Learning},
  author={Reyes, Jonatan and Di Jorio, Lisa and Low-Kam, Cecile and Kersten-Oertel, Marta},
  journal={Computational Intelligence},
  volume={41},
  number={6},
  pages={e70150},
  year={2025},
  publisher={Wiley Online Library}
}

@article{konevcny2016strategies,
  title={Federated learning: Strategies for improving communication efficiency},
  author={Kone{\v{c}}n{\`y}, Jakub and McMahan, H Brendan and Yu, Felix X and Richt{\'a}rik, Peter and Suresh, Ananda Theertha and Bacon, Dave},
  journal={arXiv preprint arXiv:1610.05492},
  year={2016}
}

@inproceedings{li2022federated,
  title={Federated learning on non-iid data silos: An experimental study},
  author={Li, Qinbin and Diao, Yiqun and Chen, Quan and He, Bingsheng},
  booktitle={2022 IEEE 38th international conference on data engineering (ICDE)},
  pages={965--978},
  year={2022},
  organization={IEEE}
}

@article{vepakomma2018split,
  title={Split learning for health: Distributed deep learning without sharing raw patient data},
  author={Vepakomma, Praneeth and Gupta, Otkrist and Swedish, Tristan and Raskar, Ramesh},
  journal={arXiv preprint arXiv:1812.00564},
  year={2018}
}

@inproceedings{he2019model,
  title={Model inversion attacks against collaborative inference},
  author={He, Zecheng and Zhang, Tianwei and Lee, Ruby B},
  booktitle={Proceedings of the 35th annual computer security applications conference},
  pages={148--162},
  year={2019}
}

@article{cihangiroglu2026lightsplit,
  title={LightSplit: Practical Privacy-Preserving Split Learning via Orthogonal Projections},
  author={Cihangiroglu, Mert and Pegoraro, Alessandro and Rieger, Phillip and Nocera, Antonino and Sadeghi, Ahmad-Reza},
  journal={arXiv preprint arXiv:2605.13265},
  year={2026}
}

@article{yang2019federated,
  title={Federated machine learning: Concept and applications},
  author={Yang, Qiang and Liu, Yang and Chen, Tianjian and Tong, Yongxin},
  journal={ACM Transactions on Intelligent Systems and Technology (TIST)},
  volume={10},
  number={2},
  pages={1--19},
  year={2019},
  publisher={ACM New York, NY, USA}
}
}

\end{document}